\documentclass[aps,pra,superscriptaddress,preprintnumbers,amsmath,amssymb]{revtex4-2}

\usepackage{braket}
\usepackage{color}
\usepackage{mathtools}
\usepackage{physics}
\usepackage[T1]{fontenc}

\usepackage[markup=none,defaultcolor=red]{changes}
\usepackage[dvipsnames]{xcolor}
\makeatletter
\AddToHook{cmd/added/before}{\def\Changes@AuthorColor{red}}
\AddToHook{cmd/deleted/before}{\def\Changes@AuthorColor{green}}
\AddToHook{cmd/replaced/before}{\def\Changes@AuthorColor{blue}}
\makeatother

\usepackage[normalem]{ulem}
\usepackage{upgreek}
\usepackage{threeparttable} 
\usepackage{multirow}
\usepackage{siunitx}

\newcommand{\beq}{\begin{equation}}
\newcommand{\eeq}{\end{equation}}
\newcommand{\beqa}{\begin{eqnarray}}
\newcommand{\eeqa}{\end{eqnarray}}
\usepackage{float}
\usepackage{booktabs}

\usepackage[colorlinks=true, linkcolor=black, citecolor=blue, urlcolor=black]{hyperref} 

\usepackage{titletoc}

\begin{document}

\title{Demonstration of quantum error detection in a silicon quantum processor}
\author{Chunhui Zhang}
\altaffiliation{These authors contributed equally to this work.}
\affiliation{Shenzhen Institute for Quantum Science and Engineering, Southern University of Science and Technology, Shenzhen 518055, China}
\affiliation{International Quantum Academy, Shenzhen 518048, China}
\author{Chunhui Li}
\altaffiliation{These authors contributed equally to this work.}
\affiliation{Shenzhen Institute for Quantum Science and Engineering, Southern University of Science and Technology, Shenzhen 518055, China}
\affiliation{International Quantum Academy, Shenzhen 518048, China}
\author{Zhen Tian}
\altaffiliation{These authors contributed equally to this work.}
\affiliation{International Quantum Academy, Shenzhen 518048, China}
\affiliation{Shenzhen Branch, Hefei National Laboratory, Shenzhen, 518048, China.}
\author{Yan Jiang}
\affiliation{Shenzhen Institute for Quantum Science and Engineering, Southern University of Science and Technology, Shenzhen 518055, China}
\affiliation{International Quantum Academy, Shenzhen 518048, China}
\author{Feng Xu}
\affiliation{Shenzhen Institute for Quantum Science and Engineering, Southern University of Science and Technology, Shenzhen 518055, China}
\affiliation{International Quantum Academy, Shenzhen 518048, China}
\affiliation{Shenzhen Branch, Hefei National Laboratory, Shenzhen, 518048, China.}
\author{Shihang Zhang}
\affiliation{Shenzhen Institute for Quantum Science and Engineering, Southern University of Science and Technology, Shenzhen 518055, China}
\affiliation{International Quantum Academy, Shenzhen 518048, China}
\author{Hao Wang}
\affiliation{Shenzhen Institute for Quantum Science and Engineering, Southern University of Science and Technology, Shenzhen 518055, China}
\affiliation{International Quantum Academy, Shenzhen 518048, China}
\author{Yu-Ning Zhang}
\affiliation{Shenzhen Institute for Quantum Science and Engineering, Southern University of Science and Technology, Shenzhen 518055, China}
\affiliation{International Quantum Academy, Shenzhen 518048, China}
\affiliation{Shenzhen Branch, Hefei National Laboratory, Shenzhen, 518048, China.}
\author{Xuesong Bai}
\affiliation{Shenzhen Institute for Quantum Science and Engineering, Southern University of Science and Technology, Shenzhen 518055, China}
\affiliation{International Quantum Academy, Shenzhen 518048, China}
\affiliation{Shenzhen Branch, Hefei National Laboratory, Shenzhen, 518048, China.}
\author{Baolong Zhao}
\affiliation{Shenzhen Institute for Quantum Science and Engineering, Southern University of Science and Technology, Shenzhen 518055, China}
\affiliation{International Quantum Academy, Shenzhen 518048, China}
\author{Yi-Fei Zhang}
\affiliation{Shenzhen Institute for Quantum Science and Engineering, Southern University of Science and Technology, Shenzhen 518055, China}
\affiliation{International Quantum Academy, Shenzhen 518048, China}
\author{Huan Shu}
\affiliation{International Quantum Academy, Shenzhen 518048, China}
\author{Jiaze Liu}
\affiliation{Shenzhen Institute for Quantum Science and Engineering, Southern University of Science and Technology, Shenzhen 518055, China}
\affiliation{International Quantum Academy, Shenzhen 518048, China}
\author{Kunrong Wu}
\affiliation{Shenzhen Institute for Quantum Science and Engineering, Southern University of Science and Technology, Shenzhen 518055, China}
\affiliation{International Quantum Academy, Shenzhen 518048, China}
\author{Chao Huang}
\affiliation{Shenzhen Institute for Quantum Science and Engineering, Southern University of Science and Technology, Shenzhen 518055, China}
\affiliation{International Quantum Academy, Shenzhen 518048, China}
\author{Keji Shi}
\affiliation{International Quantum Academy, Shenzhen 518048, China}
\author{Mingchao Duan}
\affiliation{International Quantum Academy, Shenzhen 518048, China}
\author{Tao Xin}
\affiliation{International Quantum Academy, Shenzhen 518048, China}
\author{Peihao Huang}
\affiliation{International Quantum Academy, Shenzhen 518048, China}
\author{Tianluo Pan}
\affiliation{International Quantum Academy, Shenzhen 518048, China}
\affiliation{Shenzhen Branch, Hefei National Laboratory, Shenzhen, 518048, China.}
\author{Song Liu}
\affiliation{International Quantum Academy, Shenzhen 518048, China}
\affiliation{Shenzhen Branch, Hefei National Laboratory, Shenzhen, 518048, China.}
\author{Guanyong Wang}
\email{wangguanyong@iqasz.cn}
\affiliation{International Quantum Academy, Shenzhen 518048, China}
\author{Guangchong Hu}
\email{hugc@iqasz.cn}
\affiliation{International Quantum Academy, Shenzhen 518048, China}
\affiliation{Shenzhen Branch, Hefei National Laboratory, Shenzhen, 518048, China.}
\author{Yu He}
\email{hey@iqasz.cn}
\affiliation{International Quantum Academy, Shenzhen 518048, China}
\affiliation{Shenzhen Branch, Hefei National Laboratory, Shenzhen, 518048, China.}
\author{Dapeng Yu}
\email{yudapeng@iqasz.cn}
\affiliation{International Quantum Academy, Shenzhen 518048, China}
\affiliation{Shenzhen Branch, Hefei National Laboratory, Shenzhen, 518048, China.}

\date{\today}

\begin{abstract}
Quantum error detection is essential in realizing large-scale universal quantum computation, especially for quantum error correction (QEC). However, key elements for FTQC have yet to be realized in silicon qubits. Here, we demonstrate quantum error detection on a donor-based silicon quantum processor comprising four-nuclear spin qubits and one electron spin as an auxiliary qubit. The entanglement capability of this system is validated through the establishment of two-qubit Bell state entanglement between the nuclear spins and the generation of a four-qubit Greenberger–Horne–Zeilinger (GHZ) state, achieving a GHZ state fidelity of 88.5(2.3)\%. Furthermore, by executing a four-qubit error detection circuit with the stabilizers, we successfully detect arbitrary single-qubit errors. The encoded Bell state entanglement information is recovered by performing the Pauli-frame update (PFU) via postprocessing. Based on the detected errors, we identify strongly biased noise in our system. Our results mark a significant advance toward FTQC in silicon spin qubits.

\end{abstract}

\maketitle
\section{Introduction}
\label{intro}

Silicon spin qubits~\cite{RevModPhys.95.025003, RevModPhys.85.961} are emerging as a promising platform alongside others, such as superconducting circuits, photonic circuits, trapped ions, neutral atom arrays, and nitrogen-vacancy centers. Significant milestones have been made recently in silicon quantum computing, including high-fidelity readout~\cite{zheng_rapid_2019,urdampilleta_gate_based_2019,West2019-ox}, fault-tolerant quantum gates~\cite{Noiri2022, Xue2022Quantum, mkadzik2022precision, Mills2022, Huang2024-fz, Tanttu2024-nu, Wang2024OperatingSQ, Thorvaldson:2024cww, steinacker2024300mm,Weinstein2023-ux}, integrated prototype demonstrations~\cite{Philips2022, Wang2024OperatingSQ, Weinstein2023-ux}, and qubit operation above one Kelvin~\cite{Petit2020Universal,Yang2020Operation,Camenzind2022, Huang2024-fz}. Given its compatibility with CMOS fabrication, silicon quantum computing could leverage industrial integration techniques for scaling up~\cite{Neyens2024-yo, steinacker2024300mm, stuyck2024cmos, Vinet2021-ha}.

\begin{figure*}
    \begin{center}
    \includegraphics[width=1\columnwidth]{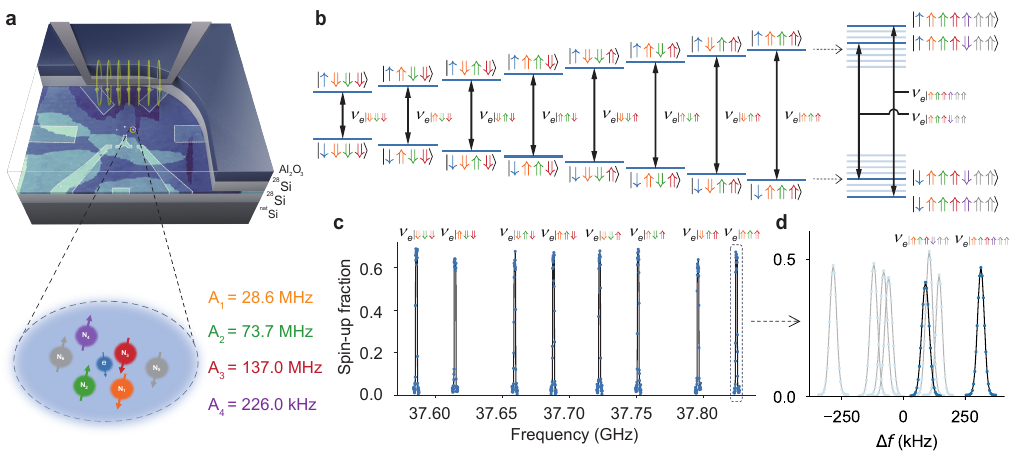}
    \end{center}
    \vspace{-0.50cm}
    \caption{\textbf{The atomic quantum processor in silicon.} \textbf{a,} Schematic illustration of the experimental device, fabricated using STM hydrogen lithography. The device contains three donor dots and here we used the right one in this experiment, which consists of five $^{31}$P nuclei and one $^1$H nucleus coupled to a shared electron through hyperfine interactions. In this study, four $^{31}$P nuclear spin qubits and one electron spin qubit are investigated, while the other nuclear spins are initialized. \textbf{b,} Energy levels and corresponding ESR transitions conditioned on specific nuclear spin configurations. \textbf{c,} ESR spectrum at a magnetic field of 1.35 T. The left panel displays eight major peaks generated by three nuclear spins: N$_1$, N$_2$, and N$_3$, with each peak comprising eight packed ESR transitions based on different spin combinations of N$_4$, N$_5$, and N$_6$. \textbf{d,} A zoom-in of the dashed box major peak on the left, conditional on N$_1$, N$_2$, and N$_3$ being on spin up $\ket{\Uparrow\Uparrow\Uparrow}$. A total of 64 ESR transition frequencies are observed. In the experiment, only two transitions labeled with black lines in the inset figure are utilized with fixed N$_5$ and N$_6$ states $\ket{\Uparrow\Uparrow}$.}
    \vspace{-0.50cm}
    \label{fig:SiliconFig1}
\end{figure*}
However, errors are inevitable in physical qubits, and the best physical qubit error rates fall short of the levels required for practical quantum computation, where $10^{-10}$ is generally needed~\cite{Acharya2024-wf}. Specifically, in silicon spin qubits, several major noise sources inherent to solid-state systems aggravate this challenge~\cite{RevModPhys.95.025003}: the embedded and interfacial defects, the surrounding flopping non-zero nuclear spins, and the lattice vibrations of the hosting materials. Collectively, these noise sources render spin qubits error-prone. Additionally, the dense arrangement of electrical gates could lead to inadvertent operation of idle qubits while controlling nearby target qubits, resulting in crosstalk errors~\cite{Tanttu2024-nu} that must be addressed as the system scales. Hence, quantum error detection and correction are highly demanded in silicon systems to bridge the gap between physical implementations and FTQC.

Detecting and correcting errors in logical states encoded across multiple physical qubits are central topics in FTQC~\cite{shor1995, steane1996multiple, preskill1998reliable, nielsen2010quantum, gottesman2010introduction}. Although phase error correction has been shown with three-~\cite{Takeda2022-sf} and four-spin qubits~\cite{Van_Riggelen2022-qq}, silicon systems have yet to demonstrate error detection with stabilizers directly. In scalable fault-tolerant error correction codes, errors that occur in the encoded logic states must be detected by the stabilizers in a quantum non-demolition manner, as required in surface code~\cite{Acharya2024-wf}, color code~\cite{Postler2022-tz}, and honeycomb code~\cite{Hastings2021dynamically}. Here, we demonstrate error detection using a four-qubit circuit in a silicon donor processor in a [[2,0,2]] code~\cite{Corcoles2015-ot}. Although the absence of a logical qubit precludes quantum error correction, we encode a Bell codeword with two qubits and employ the other two ancillary qubits to facilitate independent and simultaneous detection of phase and bit-flip errors using $\hat{S}^{X}=XX$ and $\hat{S}^{Z}=ZZ$ stabilizers, respectively. We further recover the information of the encoded entanglement state via postprocessing~\cite{Bultink2020, Andersen2019-hf}. The four-qubit error detection represents a significant advance toward FTQC in silicon systems.

\section{System}

In this experiment, we utilized a donor cluster consisting of multi-nuclear spins and one electron spin to implement the quantum circuits. The donor atom quantum processor (see Fig. 1a) is fabricated using the scanning tunneling microscope (STM) hydrogen lithography~\cite{Bussmann2021-ax, Thorvaldson:2024cww, He2019, Kranz2020, donnelly2024noise} on an in-situ grown $^{28}$Si substrate, featuring a low non-zero nuclear spin concentration of 130 ppm. The patterned device is incorporated with phosphorus atoms and includes a single-electron transistor (SET) for spin readout, three quantum dots composed of clusters of phosphorus atoms, and in-plane gates to control the relative chemical potential between the SET and the dots~\cite{Thorvaldson:2024cww, He2019, Kranz2020}. In this study, one of the quantum dots is utilized to execute the quantum circuits. A stripe-line microwave antenna is fabricated on an atomic-layer-deposited $\rm{Al_2O_3}$ dielectric layer on top of the device. This antenna facilitates coherent operations of the electron and nuclear spins via electron spin resonance (ESR) and nuclear magnetic resonance (NMR) pulses by frequency multiplexing. The device is housed in a dilution refrigerator at a base temperature of 15 mK and subjected to a parallel external magnetic field of 1.35 T.

\begin{figure*}[ht]
    \begin{center}
    \includegraphics[width=1\columnwidth]{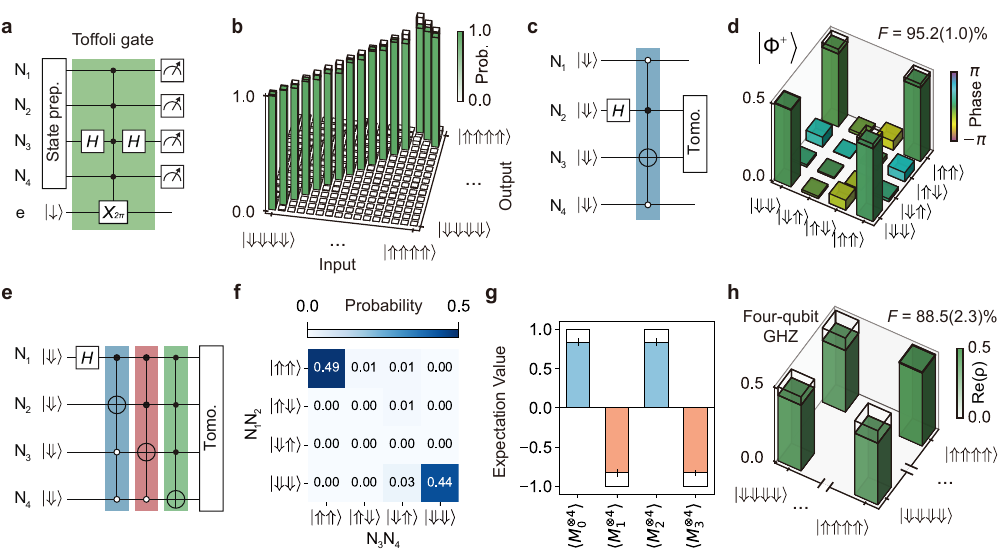}
    \end{center}
    \vspace{-0.50cm}
    \caption{\textbf{The implementation of Toffoli gate, preparation of Bell states and the four-qubit GHZ state.} \textbf{a,} Quantum circuit diagram employed to estimate the population transfer fidelity of the four-qubit Toffoli gate. \textbf{b,} The measurement results of the four-qubit Toffoli gate truth table, which gives a population transfer fidelity of 95.9(0.03)$\%$. \textbf{c,} Example circuit for establishing Bell entanglement between two nuclear spins. \textbf{d,} The state tomography of Bell state $\ket{\Phi^+}$ between N$_2$ and N$_3$. \textbf{e,} Quantum circuit diagram for generating the four-qubit GHZ state. Here we use the witness operator $W_G$ to evaluate the state preparation fidelity, $F=0.5- W_G$, where $W_G$ is determined via measurements in different bases as shown in \textbf{f} and \textbf{g}. \textbf{f,} Results from the joint measurement in the $ZZZZ$ basis are shown in a 2D array, with the first two spins (N$_1$ and N$_2$) listed vertically and the last two spins (N$_3$ and N$_4$) listed horizontally. \textbf{g,} Expectation values obtained in coherent basis. \textbf{h,} Measured corner elements of the density matrix (real part) for the four-qubit GHZ state. In circuits \textbf{a}, \textbf{c}, and \textbf{e}, the filled (open) circles indicate that the conditional rotations are activated by the $\ket{\Uparrow}$ ($\ket{\Downarrow}$) state of the control qubits. The error bars in \textbf{g} represent 1$\sigma$ confidence intervals.}
    \vspace{-0.50cm}
    \label{fig:SiliconFig2}
\end{figure*}

Our multi-spin system comprises an electron and six nuclear spins, where the electron is bound to a cluster of five phosphorus atoms and an adjacent hydrogen nucleus. By employing a linear frequency-chirped pulse with a 1.5 MHz scan to adiabatically flip the electron spin, we discern eight ESR transitions, as shown in Fig. 1b. These transitions arise from the interaction between the electron spin and three phosphorus nuclear spins (labeled as N$_1$, N$_2$ and N$_3$, respectively) through hyperfine interaction, where $A_1$ = 28.6 MHz, $A_2$ = 73.7 MHz, and $A_3$ = 137.0 MHz. Zooming into each transition with a finer scan uncovers an additional eight closely packed transitions, shown in the right panels of Fig. 1b,  associated with two further placed phosphorus donor nuclear spins, N$_4$ and N$_5$, as well as a hydrogen nuclear spin, N$_6$. In total, we observe 64 ESR transitions, with every eight transitions grouped. In this study, we utilize spins N$_1$-N$_4$ along with the electron spin to execute circuits while initializing the other nuclear spins N$_5$ and N$_6$ to a fixed spin state to mitigate uncontrolled dynamics and crosstalk. A detailed description of the system and modeling are provided in the Methods. The Rabi frequencies of the electron and the nuclear spins are typically around 250 kHz and 10 kHz in this study, using ESR and NMR driving, respectively. Their corresponding dephasing times $T_{2}^{\ast}$ are about 23.4(0.5) $\upmu$s, 441(11) $\upmu$s, 349(8) $\upmu$s, 788(23) $\upmu$s and 24.8(0.8) ms (see Extended Data Fig. 1-2). 

The single-cluster architecture suffers from inherent scalability limitations due to frequency crowding as the number of nuclei increases, imposing an upper bound on the number of nuclear spins per cluster. To overcome this constraint, cluster arrays can be employed, leveraging switchable electron exchange interactions between neighboring sites~\cite{He2019} (see Supplementary Information). A recent experiment demonstrated full connectivity and entanglement between nuclear spins in two adjacent clusters mediated by exchange interactions~\cite{edlbauer2025}. Although large-scale donor cluster arrays presents challenges, such as electrode fan-out and frequency crosstalk, these issues can be resolved through further advances in engineering technologies.

\section{Entanglement generation}
We proceed to validate entanglement generation capabilities in our system. All quantum gates in this donor cluster are executed through the ESR and NMR operations. In addition to NMR-type single-qubit gates for nuclear spins, a native gate set is the ESR-type multi-qubit controlled nuclear CCCZ type gates. Performing a 2$\pi$ rotation on a given ESR transition results in the corresponding specific nuclear spin combination state acquiring a $\pi$ geometric phase, and thus we can effectively implement a nuclear CCCZ gate with the assistance of the electron~\cite{mkadzik2022precision, Waldherr2014-pk, Thorvaldson:2024cww}. As this CCCZ gate features intrinsic high connectivity and functions as a fast geometric gate, the Toffoli gates on the nuclear spins in our system can be easily realized by combining the CCCZ gate and nuclear single-qubit gates, as shown in Fig. 2a. Compared to the classical methods proposed to construct the three-qubit Toffoli gate, requiring at least five two-qubit gates and placing a substantial overhead for multi-qubit operations~\cite{PhysRevA.88.010304}, the Toffoli gate presented here is both efficient and straightforward to implement for generating multi-qubit GHZ entanglement state in our system. The truth table $U_{\rm exp}$ for the four-qubit Toffoli gate is depicted in Fig. 2b, based on measurements of four-qubit input and output states in the Pauli $\sigma_z$ basis. Our four-qubit Toffoli gate population transfer fidelity is $\mathrm{Tr}(U_{\rm exp}U_{\rm ideal})/16 = 95.9(0.03)\%$. 
Subsequently, Clifford gates are constructed using the native single-qubit gates and CCCZ type gates. The Clifford-based randomized benchmarking (RB) and interleaved randomized benchmarking (IRB) techniques are applied to evaluate our gate fidelities. On average, we realize a nuclear spin single-qubit Clifford gate fidelity of $F_{1Q}= 99.57(0.01)\%$ and a two-qubit CZ gate fidelity of $F_{2Q}= 97.76(0.42)\%$ across all pairs among the four encoded nuclear spins. Detailed characterizations of all individual single- and two-qubit gates can be found in Extended Data Fig. 4-5.

\begin{figure*}[ht]
    \begin{center}
    \includegraphics[width=1\columnwidth]{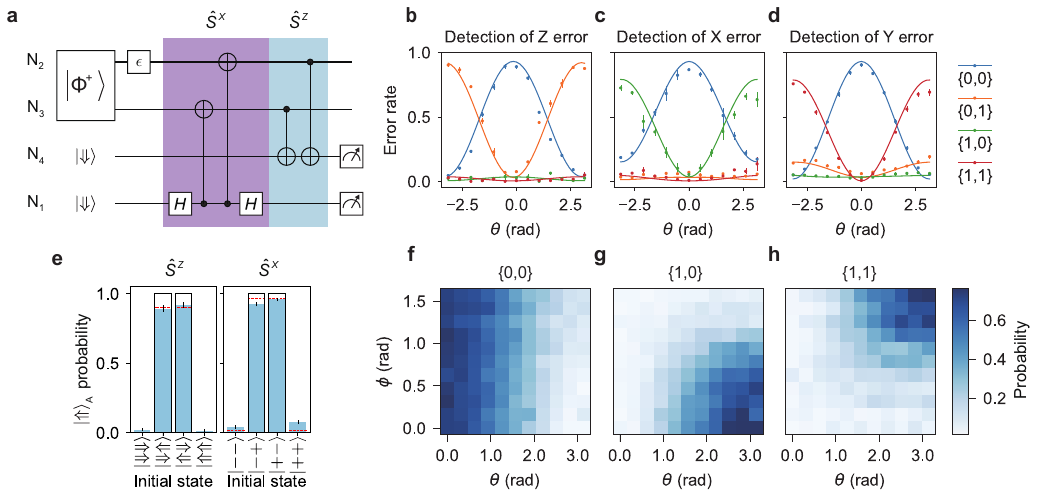}
    \end{center}
    \vspace{-0.50cm}
    \caption{\textbf{Arbitrary single-qubit error detection using stabilizer measurements.} \textbf{a,} Schematic representation of the quantum circuit for detecting arbitrary single-qubit errors. The error $\epsilon=e^{-i\theta \hat{n}\cdot\vec{\sigma}/2}$ is introduced by a single-qubit rotation, followed by the execution of the error syndrome circuit and performing parity measurements of the $\hat{S}^{Z}$ and $\hat{S}^{X}$ stabilizers. \textbf{b-d,} Error rates of detecting three different types of errors: Z, X, and Y errors, respectively. The Y error can be decomposed into both X and Z errors. The solid lines indicate the simulated results. \textbf{e,} Stabilizer values for different input states of N$_2$ and N$_3$. The left and right panels correspond to the measurement outcomes of stabilizers $\hat{S}^{Z}$ and $\hat{S}^{X}$, respectively. Basis $\ket{+}$ and $\ket{-}$ are defined as $\ket{+}\equiv\frac{1}{\sqrt{2}}(\ket{\Downarrow}+\ket{\Uparrow})$ and $\ket{-}\equiv\frac{1}{\sqrt{2}}(\ket{\Downarrow}-\ket{\Uparrow})$, respectively. The red dashed lines represent the simulated results. \textbf{f-h,} Continuous error detection as the error probability and types evolve. Here, $\epsilon=e^{-i\theta(\text{cos}\phi\cdot\sigma_{x}+\text{sin}\phi\cdot\sigma_{y})/2}$, when $\phi$=0, the error is a pure X error, and as $\phi$ increases from 0 to $\pi/2$, the X error gradually transitions into a Y error.}
    \vspace{-0.50cm}
    \label{fig:SiliconFig3}
\end{figure*}

Since this CCCZ gate provides an all-to-all connection between nuclear spins, we establish the Bell state entanglement between every pair for nuclear spin combinations using the circuits shown in Fig. 2c. The fidelities of two-qubit Bell state $\ket{\Phi^+}=\frac{1}{\sqrt{2}}(\ket{\Uparrow\Uparrow}+\ket{\Downarrow\Downarrow})$ are evaluated by full state tomography, as shown in Fig. 2d. The measured average Bell state fidelity is $F_{\rm \ket{\Phi^+}}^{\rm{avg}} = 93.4(0.5)\%$ (all the two-qubit entanglement results are presented in Extended Data Fig. 6). The entanglement among all the encoded nuclear spins is demonstrated through the generation of a four-qubit GHZ state with the circuit depicted in Fig. 2e. Despite the relatively short electron spin coherence time compared to the duration of single nuclear spin gate operations, it does not adversely affect the multi-nuclear entanglement state, as the electron spin admits faster control intrinsically and only experiences a single 2$\pi$ rotation. To assess the fidelity of the four-qubit GHZ state, we first measure it in the computational basis, confirming that only the states $\ket{\Uparrow \Uparrow \Uparrow \Uparrow}$ and $\ket{\Downarrow \Downarrow \Downarrow \Downarrow}$ are predominantly populated (Fig. 2f). We then verify that the state is indeed in coherent superposition states by measuring with the coherent basis operators $M_{0}^{\otimes4}$, $M_{1}^{\otimes4}$, $M_{2}^{\otimes4}$, and $M_{3}^{\otimes4}$, where $M_{0}=\sigma_{x}$, $M_{1}=[(\sigma_{x}+ \sigma_{y})/\sqrt{2}]$, $M_{2}=\sigma_{y}$, and $M_{3}=[(\sigma_{x}- \sigma_{y})/\sqrt{2}]$, respectively, with $\sigma_{x}$ and $\sigma_{y}$ are the Pauli matrices. The coherent basis measured expectation values are presented in Fig. 2g. Subsequently, we determine the fidelity of the four-qubit GHZ state using the entanglement witness operators~\cite{PhysRevLett.92.087902}. Based on the results shown in Fig. 2e and Fig. 2f, we obtain a GHZ state fidelity of 88.5(2.3)\% and subsequently reconstruct the real part of its density matrix extrema, presented in Fig. 2h. These results manifest genuine multipartite GHZ entanglement.

\begin{figure*}[ht]
    \begin{center}
    \includegraphics[width=1\columnwidth]{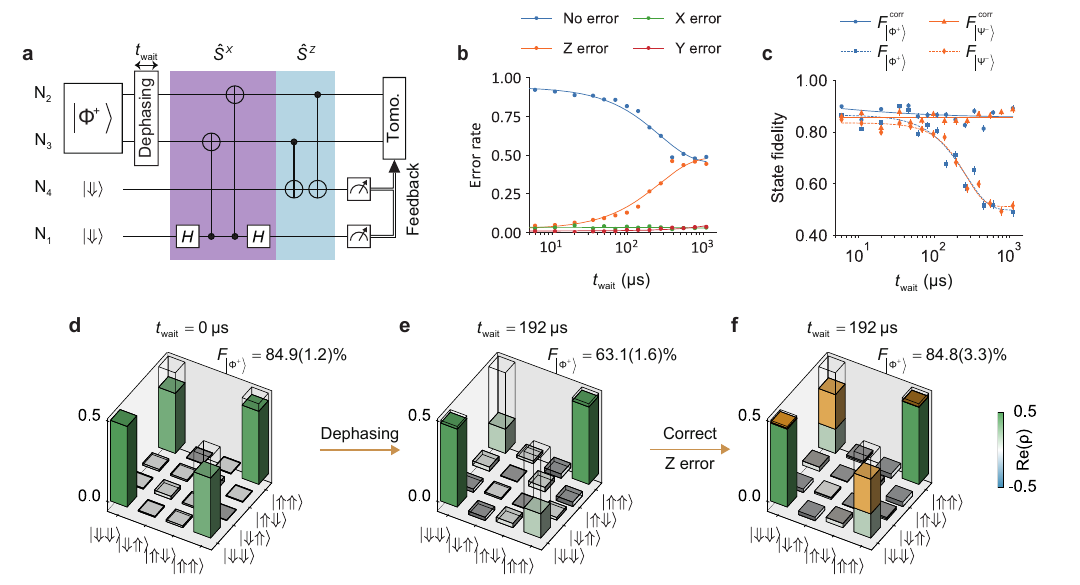}
    \end{center}
    \vspace{-0.50cm}
    \caption{\textbf{Restoration of encoded entanglement information.} \textbf{a,} Schematic representation of the quantum circuit for correcting dephasing errors. The Bell state $\ket{\Phi^+}$ undergoes a waiting time $t_\text{wait}$ prior to the stabilizer measurement. The incorrect result is postprocessed using the PFU based on the error syndrome. \textbf{b,} Measurement of the error rate of uncorrected Bell state $\ket{\Phi^+}$ for different syndromes as the dephasing time increases. \textbf{c,} Comparison of the state fidelities between the restored and uncorrected Bell codeword $\ket{\Phi^+}$ as well as the comparison for $\ket{\Psi^-}$. Z errors are corrected to obtain the postprocessed state fidelity. The dashed curves are fitted with an exponential decay function of $f(t) = A\mathrm{exp}(-(t/T_{2})^{\alpha})+B$~\cite{Bradley2019,steinacker2024violat} . Error bars are 1$\sigma$ confidence intervals. \textbf{d-f,} Density matrices for the initial $\ket{\Phi^+}$ at $t_\text{wait}$ = 0 $\upmu$s, after dephasing at $t_\text{wait}$ = 192 $\upmu$s, and after Z  error correction at $t_\text{wait}$ = 192 $\upmu$s, and the yellow bars in Fig. 4f represent the values from the restored Bell state information.}
    \vspace{-0.50cm}
    \label{fig:SiliconFig4}
\end{figure*}

\section{Error detection}
We are now prepared to perform quantum error detection in our system. The circuit in Fig. 3a involves four qubits and represents a primitive, nontrivial cut of surface code, functioning as a [[2,0,2]] code with zero logic qubits and a distance of 2~\cite{Corcoles2015-ot}. This circuit is capable of detecting both phase-flip and bit-flip errors via the $\hat{S}^{X}$ and $\hat{S}^{Z}$ stabilizers. In the circuit, $\rm{N_2}$ and $\rm{N_3}$ act as the code qubits, prepared in a Bell codeword state, while $\rm{N_1}$ and $\rm{N_4}$ serve as ancillary syndrome qubits for $\hat{S}^{X}$ and $\hat{S}^{Z}$ stabilizer measurements, respectively (the compiled circuit can be found in Supplementary Information). For the purpose of illustrating quantum error detection, we define the syndrome qubit states as $\ket{0}\equiv\ket{\Downarrow}$ and $\ket{1}\equiv\ket{\Uparrow}$. Depending on the stabilizer measurement outcomes: $\{0,0\}$, $\{0,1\}$, $\{1,0\}$, and $\{1,1\}$, the input codeword is stabilized into four Bell state subspaces. Through these stabilizer syndromes, we can detect an arbitrary single-qubit error, and then restore the original Bell codeword information via conditional real-time feedback or PFU in postprocessing~\cite{Andersen2019-hf, Bultink2020}, and further details are supplemented in the Supplementary Information.

First, we characterize the stabilizer measurements within our system. Figure 3e presents that the $\hat{S}^{Z}$ and $\hat{S}^{X}$ stabilizers correctly assign the prepared basis states onto the syndrome measurement outcomes. We achieve success probabilities of 90.6(1.5)\% and 94.03(1.0)\%, respectively, by calculating overlaps between the measured values and the ideal values. In Fig. 3e, the black frame denotes the ideal values, while the red frame indicates results obtained from master-equation simulations, with close alignment affirming the high fidelity of the stabilizer measurements. We would like to strengthen that high-fidelity stabilizer measurement is essential for quantum error detection, avoiding the unambiguous error categorization. High-fidelity stabilizer measurement requires high-fidelity quantum gates and ancillary readout. In our system, this is enabled by the high-fidelity single-qubit gates on the nuclear spins, high-connectivity CCCZ type gates, and high-fidelity nuclear spin quantum non-demolition (QND) measurements. To test continuous error tracking, we implemented Z, X, and Y errors formulated as $\epsilon=e^{-i\theta \cdot\sigma_{i}/2}$, with $\sigma_{i}$ taking $\sigma_{z}$, $\sigma_{x}$, and $\sigma_{y}$, respectively, while $\theta$ varying between $-\pi$ and $+\pi$. The associated populations of the four syndrome qubit states are plotted in Fig. 3b-d. Take Fig. 3d for instance, which shows the continuous tracking of Y error, as the magnitude of the Y error increases from 0 to $\pi$, the majority syndrome outcome \{0,0\} changed to \{1,1\}, indicating no error transitioned into Y errors. In contrast, the other two cases of \{0,1\} and \{1,0\} reflecting Z and X errors remain with low probabilities. Similarly, successful tracking of Z and X errors are shown in Fig. 3b and Fig. 3c, respectively.

To demonstrate arbitrary error detection, we introduced an error in the form $\epsilon=e^{-i\theta(\text{cos}\phi\cdot\sigma_{x}+\text{sin}\phi\cdot\sigma_{y})/2}$. For stabilizer measurement results \{0,0\}, \{1,0\}, \{1,1\}, corresponding to no error, X error, and Y error, respective experimental results are plotted in Fig. 3f-h. Each panel shows the detection probabilities of these syndromes, with an increase in the error rate on the horizontal axis, and a transition from X error to Y error on the vertical axis. Also, pure X error exists in the bottom right corner, and pure Y error in the upper right corner. Figure 3f illustrates the results with no error, revealing a reduction in value along the X-axis, with minimal variation on the Y-axis. Figure 3g displays the X error outcomes, showing a majority distribution at the bottom right corner. Figure 3h presents the results for Y errors, exhibiting a diagonal gradient that confirms the detection of Y errors. Collectively, these results demonstrate arbitrary error detection using $\hat{S}^{X}$ and $\hat{S}^{Z}$ stabilizers in a quantum non-demolition manner.
    
Utilizing stabilizers to detect arbitrary single-qubit errors enables the preservation of entanglement information through quantum error corrections corresponding to the syndromes~\cite{Andersen2019-hf,Bultink2020}. Here, we utilize PFU to restore entanglement. We generate a uniform superposition state of the four Bell codewords $\ket{\psi} = \frac{1}{\sqrt{2}}(\ket{\Downarrow\Downarrow}+\ket{\Downarrow\Uparrow}) = \frac{1}{2}(\ket{\Phi^+} + \ket{\Phi^-} + \ket{\Psi^+} + \ket{\Psi^-})$ to prove that the $\hat{S}^{X}$ and $\hat{S}^{Z}$ stabilizers can restore the entanglement information from the error subspaces, where $\ket{\Phi^+}=\frac{1}{\sqrt{2}}(\ket{\Downarrow\Downarrow}+\ket{\Uparrow\Uparrow})$, $\ket{\Phi^-}=\frac{1}{\sqrt{2}}(\ket{\Downarrow\Downarrow}-\ket{\Uparrow\Uparrow})$, $\ket{\Psi^+}=\frac{1}{\sqrt{2}}(\ket{\Downarrow\Uparrow}+\ket{\Uparrow\Downarrow})$, and $\ket{\Psi^-}=\frac{1}{\sqrt{2}}(\ket{\Downarrow\Uparrow}-\ket{\Uparrow\Downarrow})$ are the four Bell basis. The readout of stabilizers subsequently assigns these states into different error subspaces with equal probability. After applying the corresponding PFU, we successfully recover the target Bell state with 79.0(1.4)\% fidelity, without postselection (see Extended Data Fig. 7). Despite being unable to execute real-time feedback due to our electron spin readout time exceeding the coherence time of the physical nuclear spin qubits, we note that entanglement or the encoded logic state can be repeatedly preserved~\cite{Bultink2020, Acharya2024-wf} in the future with enhanced readout speed ~\cite{PhysRevX.9.041003}.

Next, we realize intrinsic decoherence error detection, demonstrating that the entanglement information is retained beyond a single qubit dephasing time. By initializing the Bell state and varying the waiting time $t_\text{wait}$, we induce decoherence before recovering entanglement based on the outcomes of stabilizers at the end of the circuit (see Fig. 4a). In Fig. 4b, we prepare the $\ket{\Phi^+}$ state, and measure the error rates for different syndromes versus $t_\text{wait}$. In Fig. 4c, for prepared $\ket{\Phi^+}$ and $\ket{\Psi^-}$ Bell codewords, we display the state fidelity for the uncorrected and recovered state fidelities. In Fig. 4b and Fig. 4c, we observe an increase of $\hat{S}^{X}$ stabilizer detected Z error and a decline in the fidelity of the uncorrected state over time, primarily attributed to the dephasing process. We find that the antiparallel entangled state $\ket{\Psi^-}$ exhibits similar levels of decoherence compared to the parallel entangled state $\ket{\Phi^+}$, with the evaluated values of $T_{2,\ket{\Psi^-}}^{\ast} = 265(26)\ \upmu$s and $T_{2,\ket{\Phi^+}}^{\ast} = 262(52)\ \upmu$s. Moreover, the dephasing times of the Bell states, when compared to the single-qubit dephasing times of $T_{2, {\mathrm{N}_2}}^{\ast} = 349(8)\ \upmu$s and of $T_{2, {\mathrm{N}_3}}^{\ast} = 788(23)\ \upmu$s, suggests that no significant correlations in noise are observed in our device (see Methods for detailed discussion). This contrasts with the situation regarding charge noise, where spatial correlations have been identified over hundreds of nanometers in the donor system~\cite{donnelly2024noise}, and a similar behavior is also reported in the Si/SiGe systems~\cite{Yoneda2023-wy, PhysRevB.101.235133}. Our measurements of negligible noise correlations in the nuclear spins suggest that charge noise may not significantly impact the encoded nuclear spins in our donor cluster. Nevertheless, the underlying mechanisms of the nuclear spin noise correlations are still an open question and warrant further study. Aside from the charge noise, the major noise source for nuclear spins is the surrounding residual $^{29}$Si nuclear spins, which are typically expected to induce uncorrelated noise~\cite{PhysRevB.101.235133}.

Furthermore, in Fig. 4b, we observe that the $\hat{S}^{X}$ stabilizer effectively detects the main error, while the $\hat{S}^{Z}$ stabilizer measurements remain almost constant over the waiting time. This observation reinforces a key characteristic of silicon spin systems, where averaged $T_{2}^{\ast}$ of 569(12) $\upmu$s is considerably shorter than $T_1$ for the encoded spins, which varies from 100 s to 300s (see Extended Data Fig. 3). Our results using stabilizers affirm that dephasing constitutes the predominant noise source, with negligible relaxation effects, indicating that the noise is strongly biased in the donor cluster system, even in the presence of crosstalk error and gate errors. We anticipate a lower gate fault-tolerance threshold by designing appropriate error-correcting codes tailored to the biased error source~\cite{PhysRevLett.124.130501}. Performing a Z error correction in postprocessing reveals that the entanglement information is well preserved, with fidelity remaining close to the initial value of $F_{\ket{\Phi^+}}=84.9(1.2)\%$. This suggests that there is almost no leakage error in our nuclear spin cluster system, contrasting to the superconducting qubits~\cite{Bultink2020}. Figure 4e-f shows an instance of tomography results for the Z error correction at a fixed dephasing time of $t_\text{wait}=192~{\upmu}$s, using the initially prepared state $\ket{\Phi^+}$. Figure 4e and Fig. 4f display the uncorrected state and the recovered state, with fidelities of 63.1(1.6)\% and 84.8(3.3)\%, respectively. The nearly fully recovered Bell state fidelities show that the entangled information can be restored from systematic decoherence.

\section{Conclusions}
In summary, this paper presents quantum error detection utilizing stabilizers with a donor-based device in silicon. In a four-qubit circuit serving as a primitive instantiation of a surface code error detection, we demonstrate that an arbitrary single-qubit error can be detected. The information obtained from the stabilizers enables postprocessing error correction and preserves entanglement against system decoherence, highlighting the central theme of quantum error detection. Our results form a step towards achieving fault-tolerant silicon spin-based quantum computing platforms. Additionally, our results illustrate the distinct error characteristics of silicon donor spin systems compared to other platforms, including biased noise and the absence of state leakage. These unique properties call for tailored quantum-error correction codes. Future directions will involve demonstrating logic state initialization and operations with repeated error detections, as well as implementing fault-tolerant operations on distributed donor cluster arrays, which are essential for scaling the system.

\section{Acknowledgments}
We acknowledge C. Pan and L. Han for the dilution refrigerator technique support, X. Deng and Y. Lu for their insightful discussions, Y. Deng and M. Gao for their help polishing the figures, and L.W. Zhang for assistance with the references.

\begin{sloppypar}
\subsection*{Author Contributions} 
Z.T., H.W., Y.-N.Z., B.Z., H.S., J.L., K.W., K.S., M.D. and T.P. fabricated the device, with G.W., Y.H., S.L. and D.Y.'s supervision. C.Z., C.L., Y.J., F.X., X.B., Y.-F.Z., and C.H., performed the experiments and analyzed the data, with G.H. and Y.H.’s supervision. S. Z. developed tools to calculate the quantum dynamics for error detection and error correction, under P.H.'s supervision. F.X. developed and applied computational tools to calculate gate benchmarking and tomography, with T.X.'s supervision. C.L., C.Z., G.H. and Y.H. wrote the manuscript, with input from all coauthors. The manuscript was revised by all the authors. 

\subsection*{Funding}
This work was supported by the National Natural Science Foundation of China (Grants No. 92165210, 62174076, 11904157, 12275117), the Innovation Program for Quantum Science and Technology (No. 2021ZD0302300), Shenzhen Science and Technology Program (Grants No. KQTD20200820113010023), Guangdong Basic and Applied Basic Research Foundation (2022B1515020074).

\subsection*{Ethics declarations}
The authors declare no competing interests.

\subsection*{Code availability}
All the code relevant to this work is available from the corresponding author upon reasonable request.

\subsection*{Data Availability}
The data relevant to this study are available via Zenodo (https://zenodo.org/records/15348336).

\end{sloppypar}

\section{Methods} \label{methods}

\subsection*{Device fabrication}

The substrate utilized in this study is a p-type natural silicon with a resistivity of 50-100 $\Omega$·cm. The substrate first undergoes chemical wet cleaning to eliminate organic contaminants, followed by a low-temperature degassing process at 500 °C to remove surface adsorbates. Subsequent high-temperature flashing processes up to 1300 °C are applied to reconstruct the surface. A 20 nm buffer layer of isotopically purified $^{28}$Si, containing $\sim$130 ppm of residual spinful isotopes, is epitaxially grown on the reconstructed natural silicon substrate at 380 °C using a vertical electron-beam evaporator. This layer serves to decouple the qubit device from the nuclear spin bath of the natural silicon substrate. The isotopically purified buffer layer is subjected to high-temperature annealing up to 780 °C for 10 s to acquire sufficiently large terraces for atomic-precision lithography, and then temperature-programmed annealing to 330 °C, resulting in a high-quality Si(100)-2$\times$1 surface. This surface is immediately terminated with a mono-layer of atomic hydrogen, acting as an atomic-thin resist. 

The STM tip selectively removes the hydrogen resist using the low-temperature hydrogen depassivation lithography (HDL) technique at 77 K, creating patterns for the device and donor sites on the surface. The exposed regions are metalized by subsequent adsorbing PH$_3$ precursor at room temperature for 5 minutes and incorporation at 330 °C for 1 minute. The device is then encapsulated with a 20 nm capping layer of purified $^{28}$Si at 250 °C (0.7 nm/h) to ensure high-quality epitaxy growth. The as-encapsulated buried P-dopant device is electrically contacted with electron-beam-lithography patterned aluminum electrodes through silicon vias. Additionally, an aluminum microwave antenna is fabricated on a 10 nm atomic-layer-deposited $\rm Al_{2}O_{3}$ dielectric layer, which electrically isolates the device underneath. Al-electrodes and Al-antenna are deposited using electron-beam deposition equipment (PLASSYS MEB 550S) under high vacuum. 

\subsection*{The system Hamitonian}
The Hamiltonian of our multi-spin system is expressed as follows 
\begin{equation}
	H = \gamma_e \mathbf{B} \cdot \mathbf{S} - \sum_{i=1}^{6} \gamma_n^{i} \mathbf{B} \cdot \mathbf{I}_i + \sum_{i=1}^{6} A_i \mathbf{S} \cdot \mathbf{I}_i,
    \label{eq:H}
\end{equation}
where $\gamma_e$ is the electron gyromagnetic ratio, and $\gamma_n^{i}$ corresponds to the nuclear gyromagnetic ratio for the $i$-th nucleus. The vector $\mathbf{S}$ represents the electron spin operators, and $\mathbf{I}_i$ indicates the vector of spin operators for the $i$-th nucleus. The magnetic field $\mathbf{B}$ consists of a static component of $B_0=1.35$ T, provided by an external superconducting coil and aligned along the $\hat{z}$ axis, together with an alternating component $\mathbf{B_{ac}}$ primarily aligned along the $\hat{y}$ axis, generated via the on-chip antenna. The static magnetic field induces Zeeman splitting in both electron and nuclear spins, while the alternating magnetic field facilitates coherent control of both electron and nuclear spins through electron spin resonance (ESR) and nuclear magnetic resonance (NMR), respectively. $A_i$ denotes the strength of the hyperfine interaction between the electron spin and the $i$-th nuclear spin. These values are extracted from the ESR spectrum shown in Fig. 1b of the main text and are specified as: $A_1$ = 28.6 MHz, $A_2$ = 73.7 MHz, $A_3$ = 137.0 MHz, $A_4$ = 226 kHz, $A_5$ = 168 kHz, and $A_6$ = 211 kHz. 
Extended Data Fig. 8 displays the NMR spectrum of six nuclear spins when the electron is in $\ket{\downarrow}$. We determine the gyromagnetic ratios for nuclei $\text{N}_1$ to $\text{N}_5$ as $\gamma_n^{1,2,3,4,5} = 17.23$ MHz/T, corresponds to phosphorus (P) atoms, while $\gamma_n^6 = 42.57$ MHz/T corresponds to the hydrogen (H) atom for $\text{N}_6$ \cite{tiesinga2021codata}.

\subsection*{Electron spin readout}
The electron energy is split into $\ket{\uparrow}$ and $\ket{\downarrow}$ under the magnetic field, here we use the energy-selective readout method (Elzerman readout)~\protect\cite{elzerman2004single} to realize the spin-to-charge conversion, thus acquiring the electron spin orientation. The SET serves as both charge sensor and electron reservoir for spin readout. During the readout period, the donor energy of $\ket{\uparrow}$ and $\ket{\downarrow}$ straddles the fermi surface of the SET island. For $\ket{\uparrow}$ electron, it will tunnel into the SET island, accompanied by the $\ket{\downarrow}$ electron tunnel back to the donor shortly after; this tunnel event brings a blip signal in the SET reflected RF signal; otherwise, the $\ket{\downarrow}$  electron will stay in the donor without the blip signal. According to the method from ref.~\cite{morello2010single} and ref.~\cite{keith2019benchmarking}, our electron spin readout average fidelity is $F_M \sim81.45\%$.

Moreover, no matter the initial electron spin state is $\ket{\uparrow}$  or $\ket{\downarrow}$ , after the readout, the electron spin will always be in $\ket{\downarrow}$ , so we also use this process to initialize the electron spin.

\subsection*{Nuclear spin readout}
The nuclear spin readout is realized by specific ESR excitation combined with electron spin readout. It is a quantum non-demolition (QND) process~\cite{mkadzik2022precision}, which means the measurement does not affect the nuclear spin state.

Taking nuclear $\ket{\Downarrow}$ readout as an example, the electron is first initialized to $\ket{\downarrow}$ state, then we apply a series of $\pi$ pulses conditional on the target nuclear being in the $\ket{\Downarrow}$ state, followed by executing the electron spin readout. If the electron spin is found to $\ket{\uparrow}$, the target nuclear spin is likely to $\ket{\Downarrow}$. Otherwise, it is $\ket{\Uparrow}$. The whole process constitutes a nuclear spin single-shot readout. We use 40 times repeated QND measurement for each nuclear spin to improve the nuclear spin readout fidelity. The initialization and readout errors for all nuclear spins are provided in the Supplementary Information.

\subsection*{Crosstalk mitigation}\label{cross-talk mitigation}
The main challenge of our system is the crosstalk between nearby ESR transitions when performing CCCZ gates. This issue becomes severe when the hyperfine interactions are too weak, such as $A_4$ = 226 kHz and $A_5$ = 168 kHz, resulting in the nearly degenerate ESR frequencies for the corresponding nucleus. Consequently, an ESR pulse applied at the frequency of the target state inadvertently induces a rotation in the nearby off-resonant states. This crosstalk error can be mitigated by reducing the driving strength of the ESR pulses. However, to maintain high-fidelity gates, the decoherence time $T_2^*$ of the electron spin limits the ESR driving strength. Therefore, optimizing the Rabi frequency of the ESR pulses is essential to suppress the crosstalk errors and maintain high gate fidelity. To illustrate this point, we consider a simple system comprising a single nucleus and a single electron. The Hamiltonian is given by,
\begin{equation}
	H_{\rm 1P} = \gamma_{e}\mathbf{B\cdot S} - \gamma_{n}\mathbf{B\cdot I} + A_{s}\mathbf{S\cdot I},
\end{equation}
refer to the system Hamiltonian in Eq. \eqref{eq:H}. The Hamiltonian can be transformed into the rotating frame with $R(t) = e^{-i\omega_{\Downarrow}\hat{S}_{z}t}$, where $\omega_{\Downarrow}=\gamma_{e}B_{0}-A_{s}/2$ is the ESR frequency corresponding to nuclear spin state $\mid\Downarrow\rangle$ and $\hat{S}_{z}$ is the Pauli-$z$ operator for the electron spin. The transformed Hamiltonian in the basis $\left\{\mid\downarrow\Downarrow\rangle,\mid\uparrow\Downarrow\rangle,\mid\downarrow\Uparrow\rangle,\mid\uparrow\Uparrow\rangle\right\}$ can be written as,
\begin{equation}
	H_{\rm 1P}^{R} = \begin{pmatrix}
		0 & f_{R} & 0 & 0 \\
		f_{R} & 0 & 0 & 0 \\
		0 & 0 & -\frac{A_s}{2} & f_{R} \\
		0 & 0 & f_{R} & \frac{A_s}{2}
	\end{pmatrix}.
\end{equation}
Here, $f_{R}$ denotes the Rabi frequency, corresponding to the rotation frequency of the resonant state. In contrast, the rotation frequency of the off-resonant state is given by $\sqrt{f_{R}^{2}+A_{s}^{2}}$. The Rabi frequency should be chosen such that the rotation on the off-resonant state completes an integer multiple of $4\pi$ rotation while the target state undergoes a $2\pi$ rotation:
\begin{equation}
	\frac{2\pi}{f_{R}} = \frac{k\cdot 4\pi}{\sqrt{f_{R}^{2}+A_{s}^{2}}},
\end{equation}  
where $k$ is an integer. Therefore, the Rabi frequency can be set as $f_{R,k}=A_{s}/\sqrt{4k^{2}-1}$, effectively suppressing crosstalk errors. In this work, the Rabi frequency is set to $A_{5}/\sqrt{3}$ ($k=1$).

\subsection*{Witness and fidelity of four-qubit GHZ state}
We employ the entanglement witness operator $W_G$ to analyze the entanglement properties of the prepared four-qubit GHZ state, as an entangled state yields a negative value of $\langle W_G \rangle$. The entanglement witness operator is defined as $W_G =\frac{I}{2} - \ket{ \textrm{GHZ}_4}\bra{ \textrm{GHZ}_4}$ where $\ket{ \textrm{GHZ}_4}\bra{ \textrm{GHZ}_4}$ represents the projector of the four-qubit GHZ state, which can be decomposed as:
\begin{align}
	| \textrm{GHZ}_4\rangle\langle \textrm{GHZ}_4| = &\frac{1}{2} \left( |0\rangle\langle 0|^{\otimes 4} + |1\rangle\langle 1|^{\otimes 4} \right) \nonumber\\
	&+ \frac{1}{8} \sum_{k=0}^{3} (-1)^k M_k^{\otimes 4},
\end{align}
where $M_k=\cos \theta_k \cdot \sigma_x + \sin \theta_k \cdot\sigma_y$ and $\theta_k= k\pi/4 $, so we can extract the value of $\langle W_G \rangle$ by measuring the Pauli operator at different angles in the x-y plane. The result, illustrated in Fig. 2, indicates that $\langle W_G \rangle = -0.385(0.023)$, which is significantly negative by 10 standard deviations, thereby confirming the presence of genuine four-partite entanglement. We also calculate the fidelity of the four-qubit GHZ state as $F_{ \textrm{GHZ}} = 0.5 - \langle W_G \rangle=88.5(2.3)\%$. 

\subsection*{Full state tomography on Bell states}
To reconstruct the density matrix of the prepared Bell states, it is necessary to perform measurements in nine different bases to achieve informational completeness. This is realized by performing a rotation from $\{I,-X/2,Y/2\}$ on each qubit prior to measurement in the computational basis. Benefiting from the quantum non-demolition (QND) measurements, we are able to perform high-fidelity measurements of both the spin-up and spin-down states of the nuclear spins in a single experiment. 

The measured density matrix is calculated as $\rho = \frac{1}{4} \sum_{i,j} \langle P_{ij} \rangle P_{ij}$, where $P_{ij} = P_i \otimes P_j$, and $P_i, P_j \in \{I, \sigma_x, \sigma_y, \sigma_z\}$ are Pauli operators acting on the $i$-th and $j$-th qubits, respectively. To ensure the density matrix is physically valid, we use maximum likelihood estimation (MLE)~\cite{banaszek1999maximum} to estimate the closest physical density matrix. MLE recovers quantum states by maximizing the probability of observing the measured data. For a given experimental data $\{f_j \}$, we construct the likelihood function,

\begin{equation}
    L(\rho) = \prod_j \text{Tr}(E_j \rho)^{f_j},
\end{equation}
where the measurement operator $\{E_j \}$ satisfies $\sum_j E_j = I$. During the maximization of $L(\rho)$, the estimated density matrix $\rho$ is constrained by the physical conditions $\rho\geq0$ and $\text{Tr}(\rho)=1$. Here we choose an initial density matrix $\rho^{(0)} $ and applied the $R\rho R$ algorithm~\cite{lvovsky2004iterative} to realize the repetitive iterations,

\begin{equation}
    {\hat{\rho}}^{\left(k+1\right)}=\frac{R({\hat{\rho}}^{\left(k\right)}){\hat{\rho}}^{\left(k\right)}R({\hat{\rho}}^{\left(k\right)})}{\text{Tr}(R({\hat{\rho}}^{\left(k\right)}){\hat{\rho}}^{\left(k\right)}R({\hat{\rho}}^{\left(k\right)}))},
\end{equation}
where the operator $R\left({\hat{\rho}}^{\left(k\right)}\right)=\sum_{j}\frac{f_j}{\text{Tr}\left({\hat{\rho}}^{\left(k\right)}E_j\right)E_j}$.

\subsection*{Numerical simulations based on master equation}
The numerical simulation of dynamics of P-donor spin qubits is based on the Lindblad master equation. The Lindblad master equation is utilized to calculate the time evolution of the density operator $\rho(t)$, incorporating incoherent processes:
\begin{equation}
\frac{d \rho}{dt} = \frac{i}{\hbar}\left[\rho,H\right]+\sum_{k=1}^{h}L_{k}\rho L_{k}^{\dagger}-\frac{1}{2}\left\{L_{k}^{\dagger}L_{k},\rho\right\},
\end{equation}
where $H$ is the system Hamiltonian shown in Eq. \eqref{eq:H}, $L_{k}$ are collapse operators, which describe incoherent errors. The operators $L_{k}$ can be written as
\begin{equation}
L_{k} = \sqrt{\gamma_{k}}C_{k},
\end{equation}
where $C_{k}$ are operators describing incoherent processes, while $\gamma_{k}$ are the corresponding decoherence rates. For example, the relaxation of qubits can be represented by $L_{T_{1}} = \sigma_{-}/\sqrt{T_{1}}$, where $T_{1}$ is the relaxation time and $\sigma_{-}$ is the lowering operator of the qubit. The dephasing of qubits can be given by $L_{T_{2}} = \sqrt{\gamma(t)}\sigma_{z}$,
where $\sigma_{z}$ is the Pauli-$z$ operator of the qubit, and $\gamma(t)$ is the dephasing rate depends on the noise spectrum. In a solid-state device, $1/f$-noise usually dominates the qubit dephasing. In this work, the dephasing of qubits is approximated as a Gauss decay process:
\begin{equation}
\mathrm{e}^{-i\phi(t)} = \mathrm{e}^{-i(\phi_{0}-\frac{t^{2}}{T_{\phi}^{2}})},
\end{equation}
where $\phi_{0}$ is the phase of the qubit without dephasing, $T_{\phi}^{2}$ is the pure dephasing time of the qubit. The corresponding decoherence rate can be obtained: $\gamma(t) = d\phi(t)/dt = 2t/T_{\phi}^2$.
The relaxation time $T_{1}$ and the pure dephasing time $T_{\phi}$ used in the simulation are extracted from experimental data. Here, all collapse operators are applied individually to their corresponding qubits. In this work, the Lindblad master equation is solved numerically by treating the equation in Liouville space~\cite{PhysRevB.87.165204}.

\subsection*{Noise correlation analysis}
In our cluster system, a single electron is bounded by six nuclei. The spatial proximity of the qubits can make them susceptible to similar noise resources, called correlated noise. The correlated noise induces a similar dephasing pattern across the qubits. As a result, for states similar to the Bell state $|\Phi^{+}\rangle=1/\sqrt{2}(\ket{\Downarrow\Downarrow}+\ket{\Uparrow\Uparrow})$, the qubit decoherence is enhanced due to constructive interference. Conversely, for states similar to the Bell state $|\Psi^{-}\rangle=1/\sqrt{2}(\ket{\Downarrow\Uparrow}-\ket{\Uparrow\Downarrow})$, the qubit decoherence is reduced due to destructive interference. However, the phenomenon has not been observed in our experiments, as shown in Fig. 4 in the main text. In this work, $T_{2,\ket{\Phi^+}}^*$ is slightly shorter than $T_{2,\ket{\Psi^-}}^*$, which indicates that the noises affecting $\textup{N}_2$ and $\textup{N}_3$ exhibit only weak correlation. When noises are nearly uncorrelated, the dephasing of each individual qubit accumulates independently in the entangled state, causing the coherence time of the Bell state to become shorter than that of any individual qubit. The correlation of noise between qubits can be characterized by the correlation factor~\cite{PhysRevB.101.235133},
\begin{equation}
C = \frac{T_{2,\textup{N}_2}^{*}T_{2,\textup{N}_3}^{*}}{4}\left[\left(\frac{1}{T_{2,\ket{\Phi^{+}}}^*}\right)^2-\left(\frac{1}{T_{2,\ket{\Psi^{-}}}^*}\right)^2\right],
\end{equation}
where $T_{2,\textup{N}_2}^{*}= $ 349(8) $\upmu$s and $T_{2,\textup{N}_3}^{*}= $ 788(23) $\upmu$s are the single-qubit coherence times of N$_2$ and N$_3$, respectively. $T_{2,\ket{\Phi^{+}}}^*= 262(52) $ $\upmu$s and $T_{2,\ket{\Psi^{-}}}^*=$ 265(26) $\upmu$s are the two-qubit coherence times corresponding to the Bell state $\ket{\Phi^{+}}$ and $\ket{\Psi^{-}}$, respectively. By taking those values, we estimate $C=0.023$, thus the correlation in noise is negligible.

\subsection*{Two-qubit CNOT gate by geometric phase}
A two-qubit gate between nuclear spins is implemented by the geometric phase, which is imparted by the shared electron. For example, a CNOT gate can be implemented between nuclear $\rm{N}_1$ and $\rm{N}_3$: firstly, $\rm{N}_1$ is initialized to $\ket{\Uparrow}$, $\rm{N}_3$ is prepared in a superposition state, $(\ket{\Uparrow} + \ket{\Downarrow})/\sqrt{2}$, the resulting nuclear state follows:
$$ \ket{\Uparrow} \otimes  \frac{1}{\sqrt{2}}(\ket{\Uparrow} + \ket{\Downarrow}) = \frac{1}{\sqrt{2}} (\ket{\Uparrow\Uparrow} + \ket{\Uparrow\Downarrow}), $$

with a ESR 2$\pi$ operation conditional on $\ket{\Uparrow\Downarrow}$ state, it will be endowed with a geomtric $\pi$ phase, thus giving a negative sign to the corresponding nuclear spin state:
$$ \frac{1}{\sqrt{2}} (\ket{\Uparrow\Uparrow} + \ket{\Uparrow\Downarrow}) \overset{2\pi}{\rightarrow}  
\frac{1}{\sqrt{2}} (\ket{\Uparrow\Uparrow} - \ket{\Uparrow\Downarrow}) = \ket{\Uparrow} \otimes  \frac{1}{\sqrt{2}}(\ket{\Uparrow} - \ket{\Downarrow}).$$
After that, a $\pi$/2 NMR pulse on the nuclear $\rm{N}_3$ along y axis will complete the whole CNOT operation.

\subsection*{Pauli frame update based Bell state restoration}
Based on the results shown in Fig. 4b, Z errors are identified as the dominant source of infidelity for the Bell state. Therefore, we implement Bell state restoration by correcting these Z errors. At a waiting time of $\mathrm{t_{wait}} = 192~\upmu$s, we performed quantum state tomography on the Bell state. The tomographic results were classified according to the measurement outcomes of $\hat{S}^{X}$ stabilizer: $\ket{0}$ outcomes correspond to $\rho$, indicating no error detected; $\ket{1}$ outcomes correspond to $\rho_\mathrm{Z}$, indicating the presence of a Z error. In Fig. 4f, we applied the PFU technique in post-processing to correct the detected Z errors. The corrected density matrix $\rho_{\text{corrected}}$ is constructed by:
\begin{equation}
    \rho_{\text{corrected}} =\mathrm{P(\ket{0})}\rho + \mathrm{P(\ket{1})}\sigma_z\rho_\mathrm{Z}\sigma_z
\end{equation}
where $\mathrm{P(\ket{0})}$ and $\mathrm{P(\ket{1})}$ denote the probabilities of detecting no error and Z error, respectively. $\sigma_z$ is the Pauli operator. The fidelity of $\rho_{\text{corrected}}$ is taken as the corrected Bell state fidelity.

\renewcommand{\figurename}{Extended Data Fig.}
\setcounter{figure}{0}
\newpage
\setcounter{page}{13}

\begin{figure*}[h]
    \begin{center}
    \includegraphics[width=1\columnwidth]{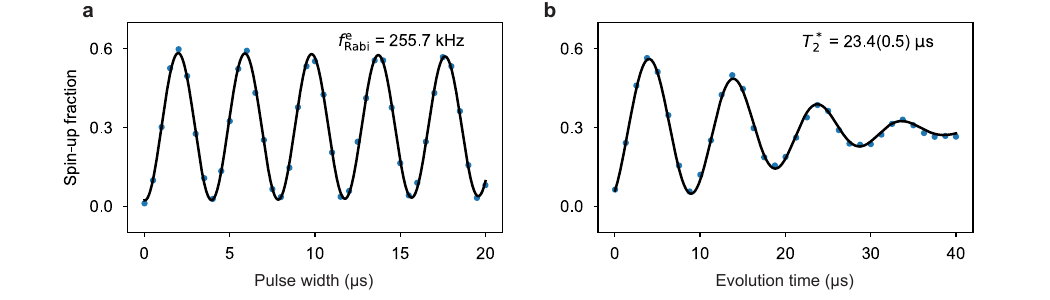}
    \end{center}
    \caption{\textbf{Rabi oscillation and Ramsey interferometry measurement for the electron.} The Rabi oscillation and Ramsey interferometry for the electron are characterized with the nuclear spins initialized in $\ket{\Downarrow\Downarrow\Downarrow\Uparrow}$ state. \textbf{a,} For Rabi measurement, a resonance ESR pulse with varying width is applied, and the extracted Rabi frequency is indicated. \textbf{b,} For Ramsey measurement, the detuning frequency is set at 100 kHz, and the integration time for each point is $\sim$128 s.}
    \label{fig:stabilizer}
\end{figure*}

\begin{figure*}[h]
    \begin{center}
    \includegraphics[width=1\columnwidth]{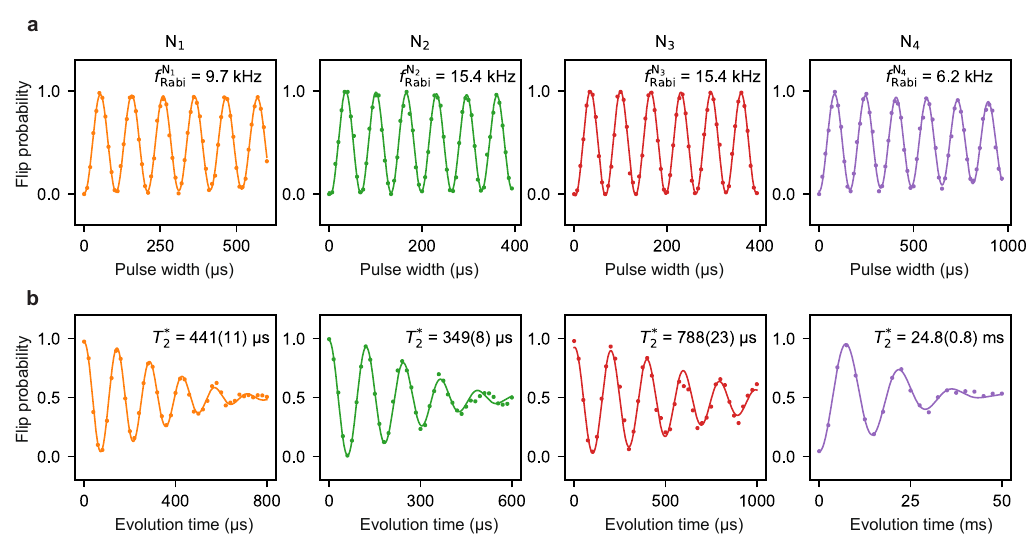}
    \end{center}
    \caption{\textbf{Rabi oscillation and Ramsey interferometry measurement for four nuclear spin qubits.} We characterize the dephasing time $T_{2}^{\ast}$ for each nuclear spin qubit using the Ramsey interferometry experiment. Here we sweep the evolution time between two $\pi$/2 pulses, and the oscillations are obtained by tuning the phase of the final $\pi$/2 pulse with $\tau$ at detunings of 7 kHz, 8 kHz, 5 kHz, and 50 Hz, respectively. The measurement sequence is repeated 40 times for each $\tau$, and 10 repetitions are performed for each nuclear spin. The equivalent integration time for $\tau$ is around 3 min for the qubits N$_{1}$N$_{2}$N$_{3}$, and 12 min for N$_{4}$. The average data is fitted by the function $P(t) = A\mathrm{exp}(-t/T_2^*)^2\cos(2\pi\Delta ft + \phi) + B$, where $\Delta f$ is the frequency detuning and $\phi$ is the initial phase, yielding dephasing times of $T_{2, \mathrm{N_1}}^{\ast}=440(11)\ \upmu$s, $T_{2, \mathrm{N_2}}^{\ast}=349(8)\ \upmu$s, $T_{2, \mathrm{N_3}}^{\ast}=788(23)\ \upmu$s, and $T_{2, \mathrm{N_4}}^{\ast}=24.8(0.8)\ $ms. The uncertainties for $T_{2}^{\ast}$ are derived by the bootstrap resampling and are denoted at the 1$\sigma$ confidence level. For $N_4$, the long dephasing time is likely due to the small hyperfine interaction of $A_{4}=226.0$ kHz.\textbf{a,} Rabi oscillation measurement results for each nucleus, performed using RF powers matching those applied in the quantum circuits of the main experiments. \textbf{b,} Ramsey interferometry measurement results for each nucleus.}
    \label{fig:stabilizer}
\end{figure*}

\begin{figure*}[h]
    \begin{center}
    \includegraphics[width=1\columnwidth]{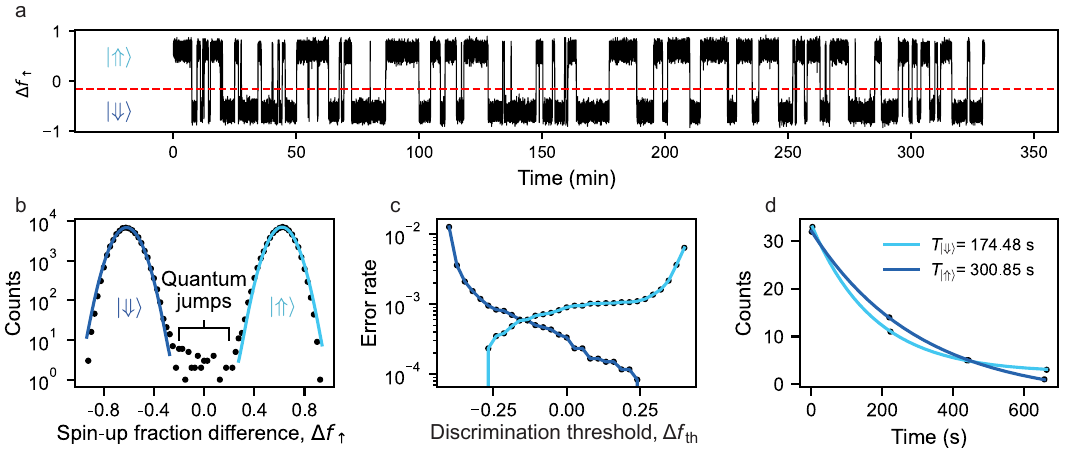}
    \end{center}
    \caption{\textbf{Nuclear spin lifetime.}\textbf{a,} Nuclear spin N$_{1}$ lifetime measurement, which is realized by toggling the two ESR frequencies corresponding to $\ket{\Uparrow}$ and $\ket{\Downarrow}$ of N$_{1}$, followed by electron spin single-shot measurement. This procedure essentially constitutes a nuclear spin quantum non-demolition measurement. Each data point represents the electron spin-up fraction difference $\Delta f_{\uparrow}$, comparing $\ket{\Uparrow}$ and $\ket{\Downarrow}$. Each spin-up fraction is averaged over 40 single-shot measurements and conducted over $\sim$80 ms. The red dotted line displays the discrimination threshold for determining whether the nuclear is in the $\ket{\Uparrow}$ or $\ket{\Downarrow}$. \textbf{b,} The quantum jumps~\cite{Pla2013} are observed throughout the measurement, with the time resolution limited by the electron spin measurement speed. In the histogram of integrated signal versus $\Delta f_{\uparrow}$, the two major peaks fitted by two Gaussian curves correspond to the two nuclear spin states. The resulting error rate for nuclear spin discrimination is drawn in \textbf{c}, giving an error rate less than 0.001 for both $\ket{\Uparrow}$ and $\ket{\Downarrow}$ states assertions, with a discrimination threshold around -0.2. \textbf{d,} The lifetimes for $\ket{\Uparrow}$ and $\ket{\Downarrow}$ of N$_{1}$ are measured over 6 hours, which is fitted using $P(t) = A\mathrm{exp}({-t/T_{1}}) + B$, yielding $\mathrm{T}^{\mathrm{N_1}}_{\ket{\Downarrow}}$ $\sim$174 s and $\mathrm{T}^{\mathrm{N_1}}_{\ket{\Uparrow}}$ $\sim$300 s. The lifetimes of nuclear spins N$_{2}$ and N$_{3}$ are characterized similarly, giving $\mathrm{T}^{\mathrm{N_2}}_{\ket{\Uparrow}}$ $\sim$100 s and $\mathrm{T}^{\mathrm{N_2}}_{\ket{\Downarrow}}$ $\sim$196 s, $\mathrm{T}^{\mathrm{N_3}}_{\ket{\Uparrow}}$ $\sim$153 s and $\mathrm{T}^{\mathrm{N_3}}_{\ket{\Downarrow}}$ $\sim$144 s, and the nuclear spin flips are mostly caused by the ionization process during electron spin readout~\cite{mkadzik2022precision}.}
    \label{fig:Lifetime }
\end{figure*}

\begin{figure*}[h]
    \begin{center}
    \includegraphics[width=1\columnwidth]{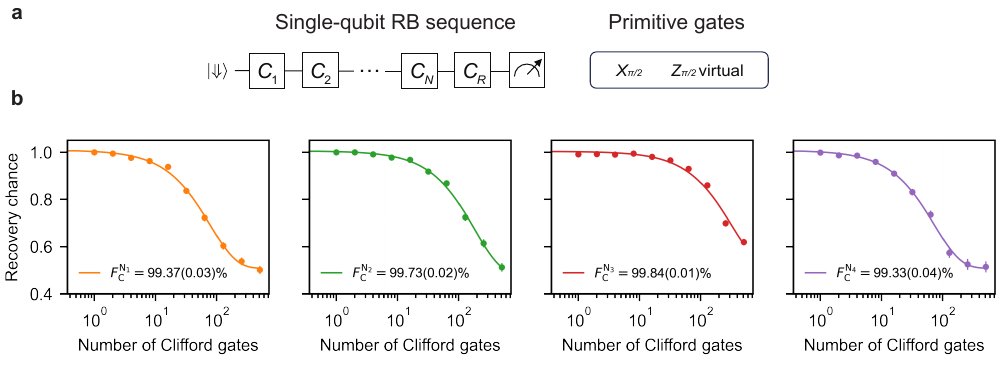}
    \end{center}
    \caption{\textbf{Single-qubit gate Randomized Benchmarking.} We characterize the single-qubit gate fidelity for four nuclear spins N$_1$-N$_4$ by performing a Randomized Benchmarking (RB) experiment. First, nuclear spins are initialized to the $\ket{\Downarrow\Downarrow\Downarrow\Downarrow\Uparrow\Uparrow}$ state. For each qubit we apply a sequence of n Clifford gates which are randomly chosen from the single-qubit Clifford group, followed by a final recovery gate to return the target qubit to its initial state. Each sequence of length $n$ is repeated 100 times, and the results are averaged over nine randomized sequences. We measure the recovery probability as a function of the number of Clifford gates, $n$, following the decay model $P_{\Downarrow}(n) = Ap_c^n  +B$ where $A$ and $B$ are constants that account for the state preparation and measurement (SPAM) errors. The depolarizing parameter $p_c$ is related to the average Clifford gate fidelity $F_C$ by $ F_C=(1+p_c)/2$. RB measurements performed on the nuclear spin qubits N$_1$-N$_4$ within their respective single-qubit subspaces yield Clifford gate fidelities of 99.37(0.03)\%, 99.73(0.02)\%, 99.84(0.01)\%,  and 99.33(0.04)\%, respectively. The fidelity uncertainties are derived by the bootstrap resampling and denoted at the 1$\sigma$ confidence level.
 \textbf{a,} The experimental circuit for single-qubit randomized benchmarking. All 24 single-qubit Clifford gates are generated using $X_{\pi/2}$ and virtual $Z_{\pi/2}$ gates. \textbf{b,} Single-qubit randomized benchmarking measurement results for four nuclear spins N$_1$-N$_4$.
    }
    \label{fig:RB}
\end{figure*}

\begin{figure*}[h]
    \begin{center}
    \includegraphics[width=1\columnwidth]{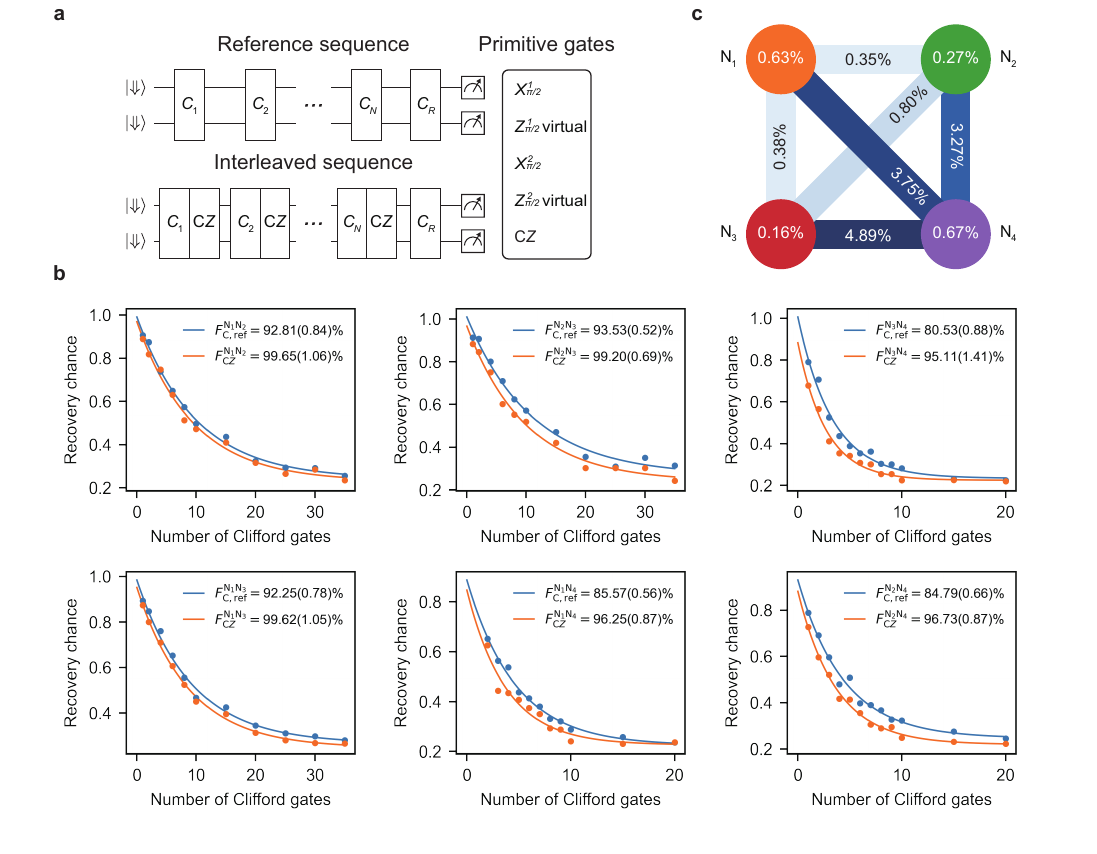}
    \end{center}
    \caption{\textbf{Two-qubit Randomized Benchmarking.} To characterize the fidelity of the CZ gate between nuclear spins $\rm{N_1}-\rm{N_4}$, we employ interleaved randomized benchmarking (IRB). All nuclear spins are initialized to the $\ket{\Downarrow\Downarrow\Downarrow\Downarrow\Uparrow\Uparrow}$ state before executing the IRB sequence. First, we perform a standard two-qubit RB experiment as a reference. Each RB sequence of length n consists of $n$ Clifford gates randomly selected from the two-qubit Clifford group, with each Clifford gate composed by an average of 4.99 $\pi/2$ pulses and 1.55 CZ gates. A recovery gate at the end of each RB sequence ensures that the two target qubits return to $\ket{\Downarrow\Downarrow}$ state. We randomly generate 9-15 distinct RB sequences from the Clifford group for each experiment, measure each sequence 200 times and average all results. By fitting the probability of $\ket{\Downarrow\Downarrow}$ by $ P_{\Downarrow\Downarrow}(n) = A p_{ref}^n + B$, we extract the reference depolarizing parameter $p_{ref}$. Subsequently, we interleave the CZ gate after each Clifford gate in the reference RB sequence, update the recovery gate, and then perform the new RB sequence to extract the depolarizing parameter $p_{CZ}$. By comparing the decay curves of the reference and interleaved sequences, the fidelity of CZ gate is extracted as $F_{CZ}= (1+3p_{CZ}/p_{ref})/4$. We characterize the two-qubit CZ gate for all pairs among the four nuclear spins, and obtain gate fidelities of 99.65(1.06)\%, 99.20(0.69)\%, 95.11(1.41)\%, 99.62(1.05)\%, 96.25(0.87)\% and 96.73(0.87)\%, respectively. The fidelity uncertainties are derived by the bootstrap resampling and are denoted at the 1$\sigma$ confidence level.\textbf{a,} The experimental circuit for two-qubit randomized benchmarking. $X_{\pi/2}^1$, $X_{\pi/2}^2$, virtual $Z_{\pi/2}^1$,  virtual $Z_{\pi/2}^2$ and CZ gate are chosen as the primitives to generate two-qubit Clifford gates.\textbf{b,} Two-qubit Randomized Benchmarking measurement results for all pairs among the four nuclear spins. The error bars represent one standard deviation calculated using the bootstrapping method. \textbf{c,} Summary of quantum gate infidelities in the four-nuclear-spin system.The values inside the circles denote the averaged infidelity of single-qubit Clifford gates for the corresponding nuclear spins, and the values between the circles represent the infidelity of the CZ gates between tem.
    }
    \label{fig:RB}
\end{figure*}

\begin{figure*}[h]
    \begin{center}
    \includegraphics[width=1\columnwidth]{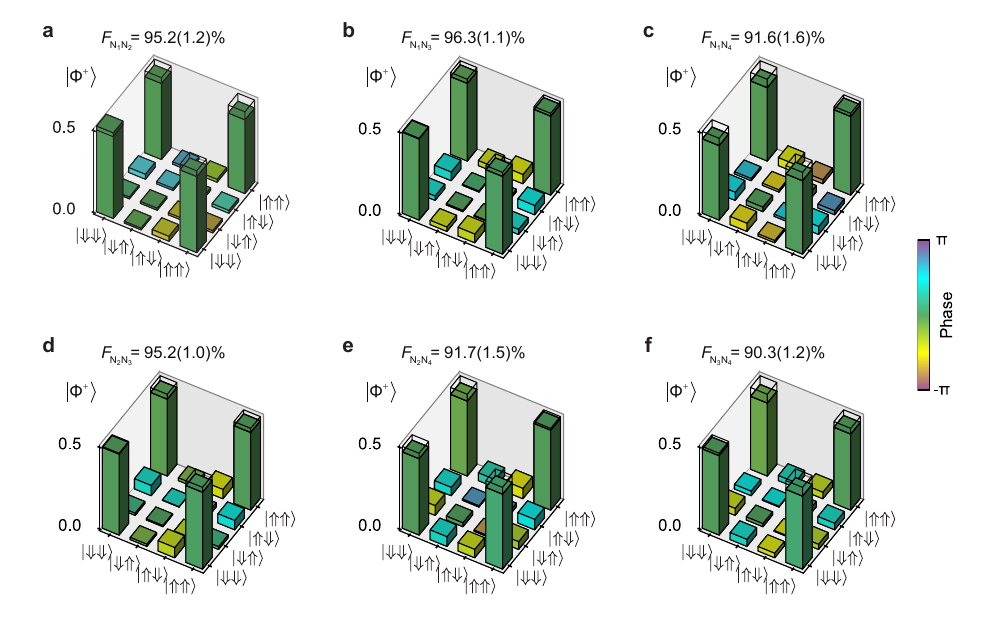}
    \end{center}
    \caption{\textbf{Bell states between each pair of the nuclear spins.} \textbf{a-f,} the tomography of Bell state $\ket{\Phi^+}$ for nuclear spin pairs N$_{1}$-N$_{2}$, N$_{1}$-N$_{3}$, N$_{1}$-N$_{4}$, N$_{2}$-N$_{3}$, N$_{2}$-N$_{4}$, and N$_{3}$-N$_{4}$, with measured Bell state fidelites of $F_{\smash{\ket{\Phi^+}},12}=95.2(1.2)\%$, $F_{\smash{\ket{\Phi^+}},13}=96.3(1.1)\%$, $F_{\smash{\ket{\Phi^+}},14}=91.6(1.6)\%$, $F_{\smash{\ket{\Phi^+}},23}=95.2(1.0)\%$, $F_{\smash{\ket{\Phi^+}},24}=91.7(1.5)\%$, and $F_{\smash{\ket{\Phi^+}},34}=90.3(1.2)\%$, respectively.}
    \label{fig:stabilizer}
\end{figure*}

\begin{figure*}[h]
    \begin{center}
    \includegraphics[width=1.0\columnwidth]{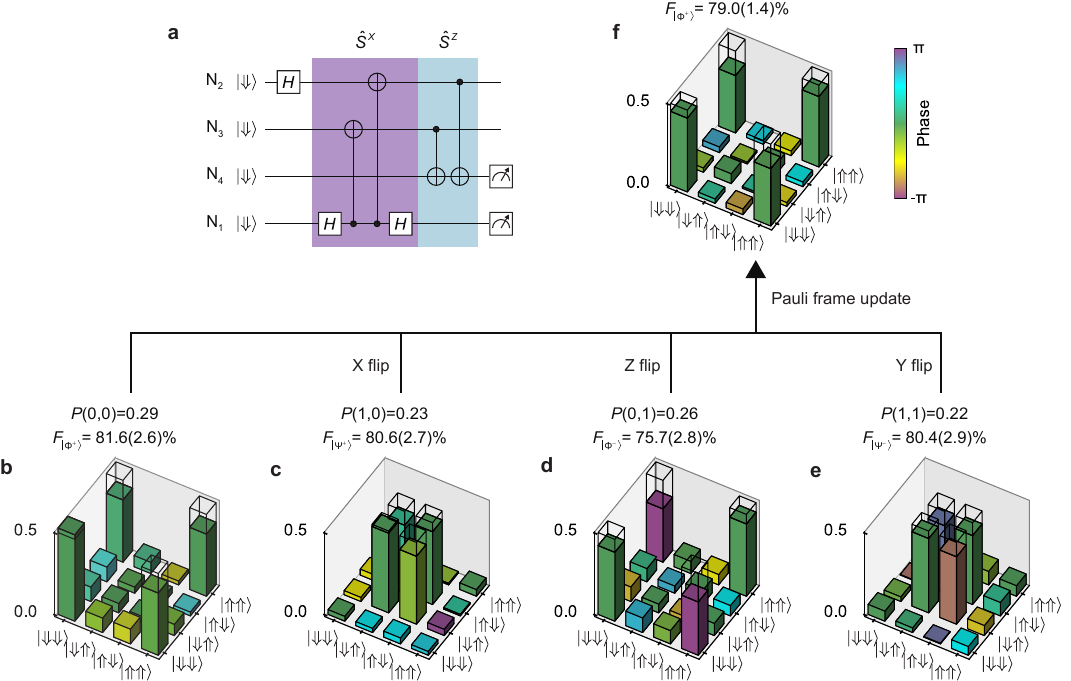}
    \end{center}
    \caption{\textbf{Bell state error correction from all the error subspace.} We demonstrate that our circuits can detect all the single-qubit errors and fully recover the prepared state from the error subspace. This capability is illustrated by preparing $\frac{1}{\sqrt{2}}(\ket{00}+\ket{01}) = \frac{1}{2}(\ket{\Phi^+} + \ket{\Phi^-} + \ket{\Psi^+} + \ket{\Psi^-})$, which is an equal superposition of the four Bell states. We then perform quantum state tomography on the output states corresponding to different error syndromes. We observe nearly equal probabilities for the four possible measurement outcomes of the syndromes: $\{0,0\}$, $\{0,1\}$, $\{1,0\}$, and $\{1,1\}$. The data qubit is mapped to corresponding Bell state with fidelities of $F_{\smash{\ket{\Phi^+}}} = 81.6(2.6)\%$, $F_{\smash{\ket{\Psi^+}}} = 80.6(2.7)\%$, $F_{\smash{\ket{\Phi^-}}} = 75.7(2.8)\%$, and $F_{\smash{\ket{\Phi^-}}} = 80.4(2.9)\%$, respectively. Based on the error syndromes, the error state $\rho_{\rm{error}}$ can be corrected by applying a corresponding single-qubit operation $U$ through $\rho_{\text{correct}} = U\rho_{\rm{error}} U^{\dagger}$ by postprocessing. As shown in the figure, by combining those operations, we could recover the target state with a fidelity of $F_{\smash{\ket{\Phi^+}}} = 79.0(1.4)\%$, without postselecting.}
    \label{fig:Bell}
\end{figure*}

\section*{Supplementary information for "demonstration of quantum error detection in a silicon quantum processor"}

\titlecontents{subsection}
  [3.5em]
  {\addvspace{6pt}}
  {\contentslabel{2.5em}}
  {}
  {\titlerule*[1em]{.}\contentspage}
  [\addvspace{2pt}]

\titlecontents{subsubsection}
  [5.5em]
  {\addvspace{4pt}\small}
  {\contentslabel{2.5em}}
  {}
  {\titlerule*[1em]{.}\contentspage}
  [\addvspace{2pt}]

\renewcommand{\thesubsection}{\Roman{subsection}}
\renewcommand{\figurename}{\textbf{Fig.}}
\renewcommand{\thefigure}{\textbf{S\arabic{figure}}}
\setcounter{figure}{0}
\setcounter{table}{0}

\setlength {\marginparwidth }{2cm} 

\startlist[local]{toc}
\printlist[local]{toc}{}{}

\newpage

\subsection{Device Details}\label{device_details}
An STM image of the hydrogen-terminated Si(100)-2×1 surface is shown in Supplementary Fig. \ref{fig:device}a, revealing large-scale atomically flat terraces. The STM tip selectively removes the hydrogen resist using low-temperature hydrogen depassivation lithography at 77 K, defining patterns for the device. Supplementary Fig. \ref{fig:device}b shows the central region of the device studied in the main text, where bright regions indicate lithographed structures which are later doped to a metallic density. The device consists of three small quantum dots formed by phosphorus clusters (Supplementary Fig. \ref{fig:device}c) with each dot containing 2--6 phosphorus atoms. This study focuses on the right one, weakly tunnel-coupled to the SET (17.0 nm away). The other two dots (17.3 nm and 17.6 nm to the SET) were not used in this study. The distances between the nearest neighbor dots are designed to be 15-17 nm apart from center to center. For this study, we fabricated three similar devices; however, the other two did not exhibit optimal SET signals for electron spin readout and were not used.

\begin{figure*}[h]
\begin{center}
    \includegraphics[width=1.0\columnwidth]{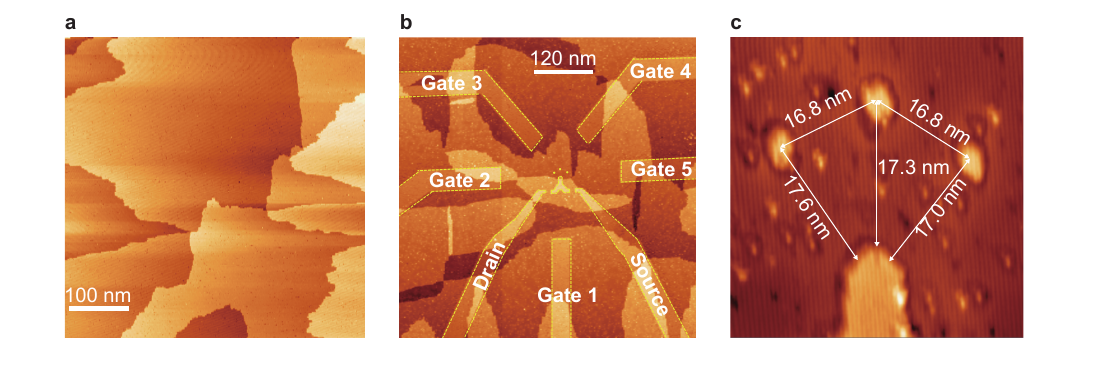}
    \end{center}
    \caption{\textbf{Device fabrication.} \textbf{a,} STM image of in situ epitaxially grown $^{28}$Si surface. \textbf{b,} STM image of the device with bright regions indicating exposed bare silicon surface, which is subsequently dosed with phosphine to create conductive leads to fanout the central region of the device. \textbf{c,} A zoom-in of the white dashed box in \textbf{b,} illustrating the central region of the device, where the innermost components consist of the single-electron transistor (SET) island and three few-donor quantum dots, with labeled distances.}
    \label{fig:device}
\end{figure*}

\begin{figure*}[h]
\begin{center}
    \includegraphics[width=0.4\columnwidth]{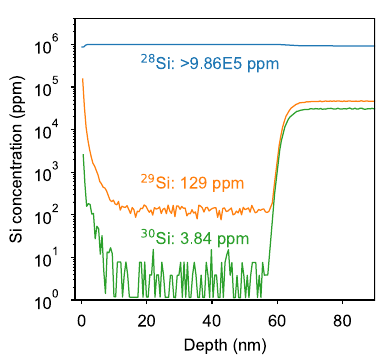}
    \end{center}
    \caption{\textbf{Secondary ion mass spectrometry.} Secondary ion mass spectrometry (SIMS) measurements are conducted to assess silicon isotope concentrations in a reference sample, which is grown under the same conditions as our device, with a thickness of approximately 50 nm. The SIMS data confirms that the concentration of the $^{29}$Si impurity is below 130 ppm.}
    \label{fig:sims}
\end{figure*}

\subsection{Experimental set-up}\label{set_up}
\begin{figure}[H]
\begin{center}
    \includegraphics[width=0.7\columnwidth]{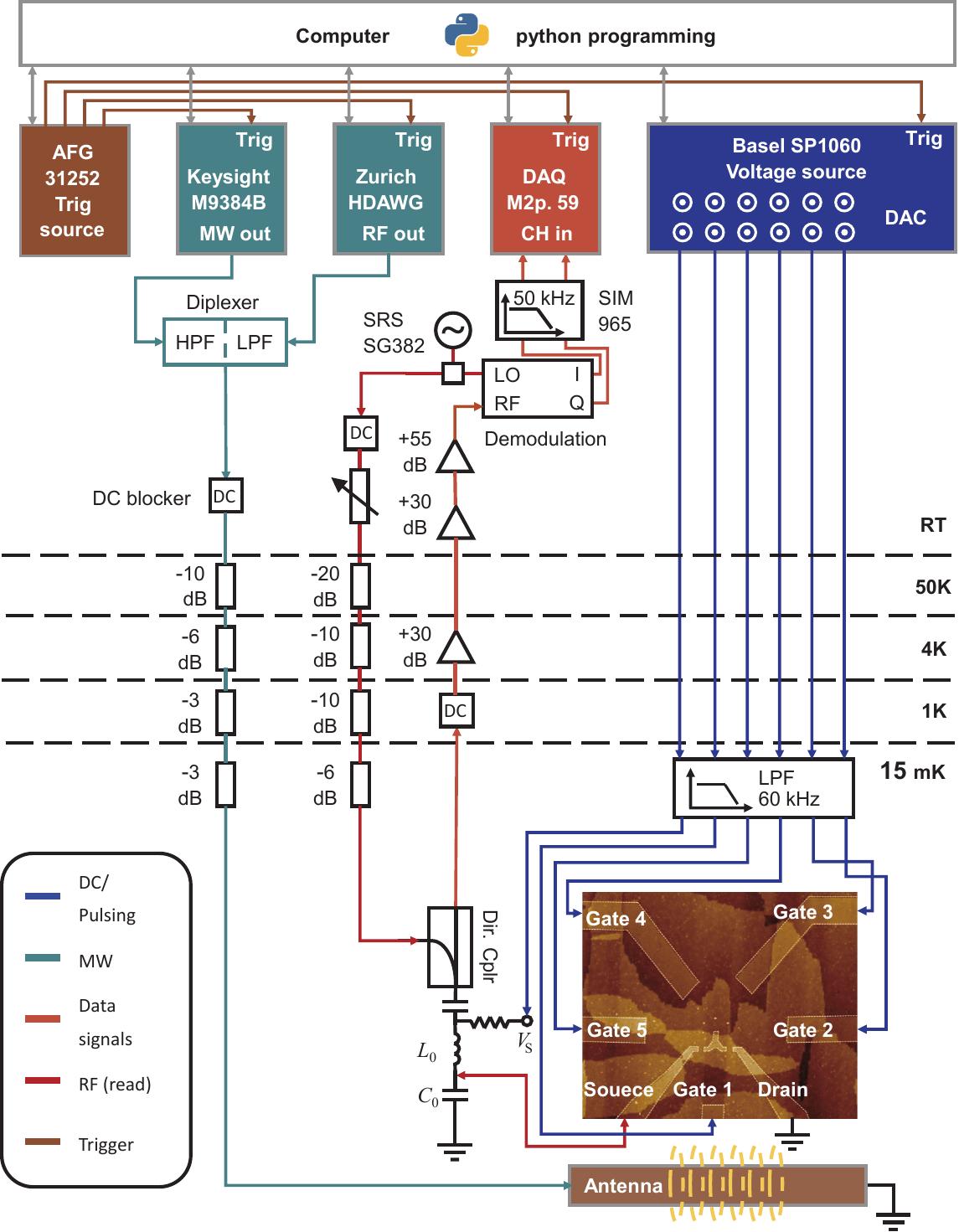}
    \end{center}
    \caption{\textbf{Experimental set-up.} All measurements are conducted in a dilution refrigerator operating at a base temperature of 15 mK and subjected to a static magnetic field of 1.35 T. DC voltages biasing the device are supplied by a 24-bit resolution, 24-channel voltage source (Basel LNHR DACII SP1060). Square voltage pulses used for electron spin readout and initialization are generated by configuring the same instrument in “AWG” mode. Each signal is filtered by a homemade low-pass RC filter placed on the mixing chamber plate, with a cutoff frequency of 60 kHz. Microwave pulses for driving the electron spin are produced by a vector signal generator (Keysight M9384B) with a baseband bandwidth of 1 GHz. The microwave carrier frequency is set at 37.707 GHz, and the output frequency is modulated to resonate with the electron spin in specific nuclear state. Radio-frequency (RF) pulses for driving the nuclear spin are generated by an arbitrary waveform generator (Zurich HDAWG). The microwave and RF signals are combined via a diplexer (Marki Microwave DPX-1721) before being directed to the broadband microwave antenna on the device. For charge sensing, we employ the radio-frequency reflectometry technique. The single electron transistor (SET) is lumped into an LC tank with a resonance frequency of 302.8 MHz. The reflected RF signal is first amplified by a commercial cryogenic amplifier, and further amplified and quadrature demodulated (Polyphase AD0105B) at room temperature. The demodulated signal is subsequently sent to a post-amplifier (SRS SIM910 JFET), combined with an analog filter (SRS SIM965) set to a cutoff frequency of 49.9 kHz. Finally, the signal is digitized by the analog-to-digital converter (M2p.5911-x4) at a sample rate of 500 kSa/s.}
    \label{fig:setup}
\end{figure}
\clearpage

\subsection{Electrostatic triangulation}\label{quantum dot}
\begin{figure*}[h]
\begin{center}
    \includegraphics[width=1.0\columnwidth]{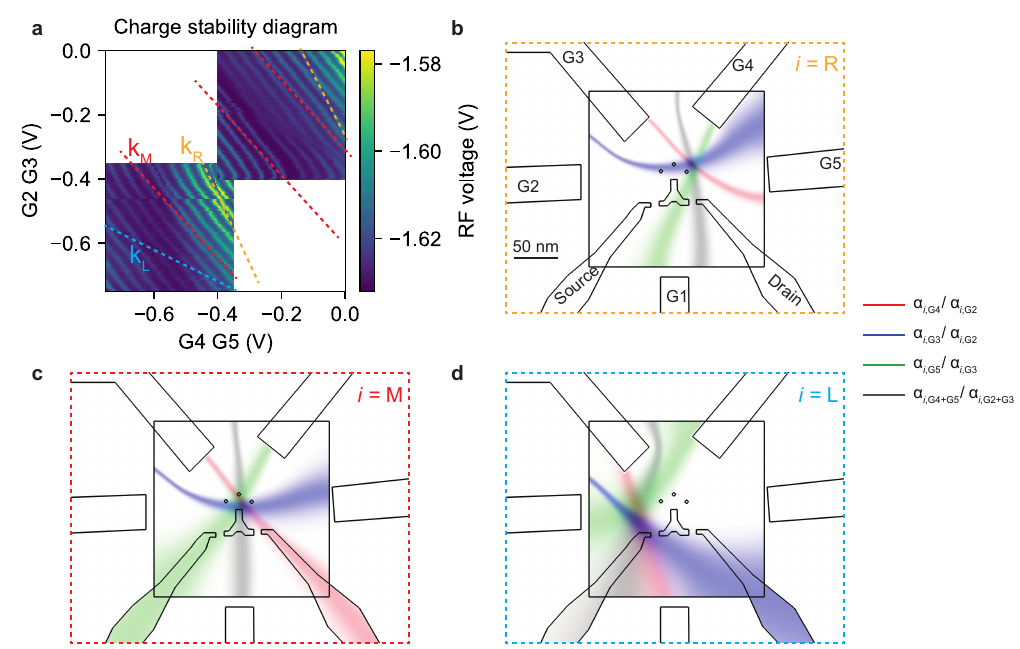}
    \end{center}
    \caption{\textbf{Electrostatic triangulations.} \textbf{a}, The charge stability diagram with a large voltage range, where three types of transition lines with different slopes can be observed. Intuitively, the yellow, red, and blue transition lines correspond to the right, middle, and left quantum dots, respectively. \textbf{b-d}, The simulation results for each transition line. Right dot in (b), middle dot in (c), and left dot in (d).}
    \label{fig:triangulation}
\end{figure*}

Here, in this section, the electrostatic triangulation results are shown to verify the locations of the STM-patterned P-donor quantum dots. By measuring the charge stability diagrams under various gate  combinations (exemplary figure shown in Fig. \ref{fig:triangulation}a in (G4+G5)/(G2+G3) gate space), we obtain charging lines with slop $k_{i,x,y} =- \alpha_{i,x}/\alpha_{i,y}$, where $\alpha_{i,x}$($\alpha_{i,y}$) is the lever-arm of gate $x$($y$) to dot $i\in\{\rm{L},\rm{M},\rm{R}\}$, and $\alpha_{i,x+x'}=\alpha_{i,x}+\alpha_{i,x'}$ if more than one gate in x-axis. By modeling the structure of the device and transferring it into an electrostatic simulation tool to simulate the electrostatic potential, we can get the spatial distribution of the lever-arm $\alpha_{x}^*(\boldsymbol{r})=V(\boldsymbol{r})|_{V_{\rm{x}}=1 \rm{V}}$ and the slop $k^*_{x,y}(\boldsymbol{r})=-\alpha_{x}^*(\boldsymbol{r})/\alpha_{y}^*(\boldsymbol{r})$.

The quantum dot position can be determined by comparing the simulated lever-arm ratio $k^*_{x,y}(\boldsymbol{r})$ with the corresponding experimentally measured slope $k_{i,x,y}$ for different gate combinations. Taking the right dot as an example, Fig. \ref{fig:triangulation}b uses the combinations G4/G2, G3/G2, G5/G3, and (G4+G5)/(G2+G3) to identify four possible location regions. Their spatial distributions $l_{i,x,y}(\boldsymbol{r})$, are visualized using a Gaussian color scheme as follows:
    \begin{equation}
        l_{i,x,y}(\boldsymbol{r}) = \exp\left\{\frac{-1}{2\sigma_k^2} \left[k^*_{x,y}(\boldsymbol{r}) - k_{i,x,y}\right]^2 \right\},
    \end{equation}
where $\sigma_k=0.1$ is the standard uncertainty in the transition line slope $k_{i,x,y}$. The intersection of four possible location regions corresponds to the simulated right quantum dot position.

\begin{table*}[h]\label{ratio}
    \centering
    
    \begin{tabular}{c|c|c|c|c}
    
    \hline
    
    \   & $\alpha_{\rm{G4}}/\alpha_{\rm{G2}}$ & $\alpha_{\rm{G3}}/\alpha_{\rm{G2}}$ & $\alpha_{\rm{G5}}/\alpha_{\rm{G3}}$   & $\alpha_{(\rm{G4}+\rm{G5})}/\alpha_{(\rm{G2}+\rm{G3})}$ \\
    
    \hline
    
    Right dot & 2.30 & 1.71 & 1.02 & 1.43  \\
    Middle dot& 1.53 & 1.47 & 0.64 & 0.99  \\
    Left dot& 0.58 & 0.93 & 0.33 & 0.48 \\

    \hline
    \end{tabular}
    \caption{\textbf{Lever-arm ratios.} \raggedright The lever-arm ratios of different gate combinations for each quantum dot. This information is measured in the charge stability diagrams of G4/G2, G3/G2, G5/G3, and (G4+G5)/(G2+G3).}
    \label{table:triangulation}
\end{table*}

Similarly, the triangulations of the middle and left quantum dots are shown in Fig. \ref{fig:triangulation}c and d. By comparing the experimental transition slopes with the triangulation result, we conclude that the utilized quantum dot is the one located on the right side of the device. The detailed transition line slope ratios are provided in the Table \ref{table:triangulation}
\clearpage

\subsection{Electron spin readout}\label{electron_spin_readout}
\begin{figure*}[h]
\begin{center}
    \includegraphics[width=1.0\columnwidth]{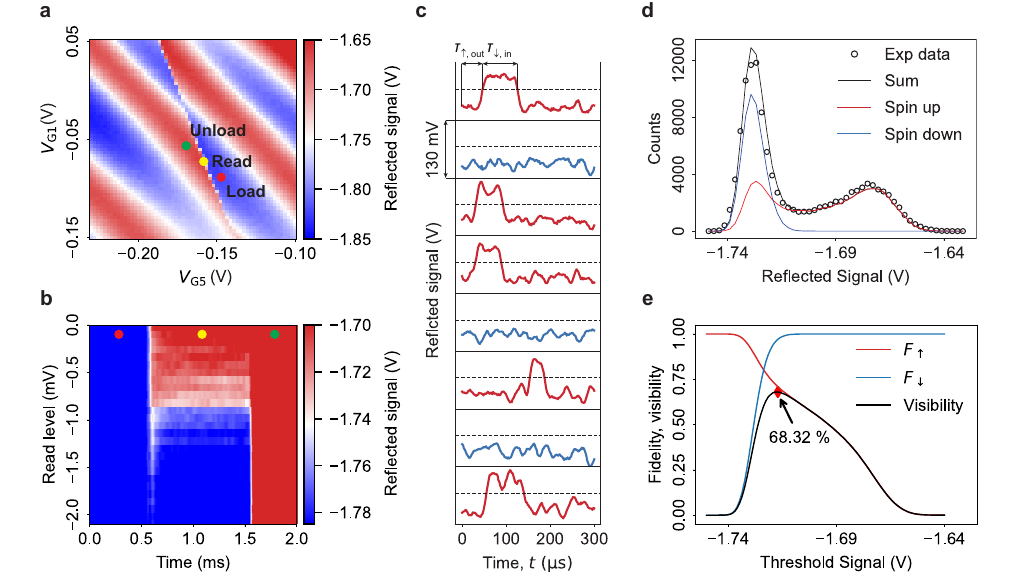}
    \end{center}
    \caption{\textbf{Electron spin readout process.} \textbf{a,} Charge stability diagram by sweeping G5 and G1 voltages. The Coulomb peaks (in red) will be broken when there is a quantum dot charge transition. To measure the spin state, a three-level pulse sequence is applied to manipulate the electron charge state, as indicated by circles ("Load" in red, "Read" in yellow and "Unload" in green). \textbf{b,} The reflected rf signal as a function of detuning voltage at the "Read" point, with a sampling bandwidth of 50 kHz, and each detuning value is averaged 1000 times. \textbf{c,}  Examples of single-shot readout traces. A blip above the threshold (line in black) indicates detection of the electron in the $\ket{\uparrow}$ state. $T_{\uparrow,\textrm{out}}$ is the tunnel-out time for an electron in the $\ket{\uparrow}$ state escaping from the cluster, while $T_{\downarrow,\textrm{in}}$ represents the subsequent tunnel-in time of an electron in the $\ket{\downarrow}$ state back to the cluster. \textbf{d,} Histogram (circles) of maximum values of reflected signals from 120,000 single-shot data traces with a readout window of 250 $\upmu$s. The solid lines are histograms of $\ket{\uparrow}$ state (red), $\ket{\downarrow}$ state (blue), and sum of both (black) using the numerical simulation method from ref.~\cite{morello2010single}. \textbf{e,} The readout fidelities of $F_E^{\uparrow}$ (red), $F_E^{\downarrow}$ (blue) and visibility $V_E$ (black) as a function of threshold, indicating a maximum visibility of 68.32\%. Further, we estimate the state-to-charge fidelities as $F_{\textrm{STC}}^{\uparrow}\sim$99.63\%, $F_{\textrm{STC}}^{\downarrow}\sim$92.56\% , and the overall measurement fidelity as $F_\mathrm{M}\sim$81.45\%.
    }
    \label{fig:spin_readout}
\end{figure*}

The degeneracy of the electron's $\ket{\uparrow}$ and $\ket{\downarrow}$ states is lifted by an external magnetic field. We use the energy-selective readout method (Elzerman readout)~\protect\cite{elzerman2004single} to realize the electron spin measurement in the $S_z$ basis. The SET serves as both a charge sensor and an electron reservoir for spin readout. Initially, the right donor dot is emptied and loaded with an electron randomly in $\ket{\uparrow}$ or $\ket{\downarrow}$ state from the reservoir. During the readout, the energy of $\ket{\uparrow}$ and $\ket{\downarrow}$ straddles the Fermi surface of the SET island. For an electron with $\ket{\uparrow}$, it will tunnel into the SET island, accompanied shortly by an electron with $\ket{\downarrow}$ tunneling back; this process results in a blip signal in the SET reflected RF signal. Otherwise, the electron with $\ket{\downarrow}$ stays on the donor without generating the blip signal, as shown in Fig. \ref{fig:spin_readout}. According to the method from ref.~\cite{morello2010single} and ref.~\cite{keith2019benchmarking}, the averaged electron spin readout fidelity is $F_\mathrm{M} \sim 81.45\%$.

Notably, after spin readout, the electron spin will be always in $\ket{\downarrow}$ regardless of its initial state. This feature is commonly used for electron spin initialization.

\clearpage

\subsection{ESR spectrum}\label{esr_spectrum}
In our experiment, we first carried out repeated ESR frequency scanning over several hours using a 10 MHz chirped pulse within the range $[\gamma_e\mathrm{B_0} - 200, \gamma_e\mathrm{B_0} + 200]$ MHz. This coarse identified eight ESR frequencies corresponding to different spin configurations of nuclei $\mathrm{N_1}$, $\mathrm{N_2}$, and $\mathrm{N_3}$. We then performed fine scans around these frequencies using a 1 MHz chirped pulse, narrowing the spectral linewidth to within 1 MHz range. Next, we calculate the hyperfine interactions $A_1$, $A_2$, and $A_3$ using the relations $A_1 = \nu_{e|\Uparrow\Downarrow\Downarrow} - \nu_{e|\Downarrow\Downarrow\Downarrow}$, $A_2 = \nu_{e|\Downarrow\Uparrow\Downarrow} - \nu_{e|\Downarrow\Downarrow\Downarrow}$, $A_3 = \nu_{e|\Downarrow\Downarrow\Uparrow} - \nu_{e|\Downarrow\Downarrow\Downarrow}$. The corresponding NMR frequencies of the nuclei conditional on electron spin-down state are given by $f_{\mathrm{N_i}} = \gamma_n\mathrm{B_0} - A_i/2$. Fine NMR spectral scans were performed around the calculated $f_{\mathrm{N_i}}$ to obtain $f_{\mathrm{N_1}}$, $f_{\mathrm{N_2}}$, and $f_{\mathrm{N_3}}$.

We employed an initialization protocol in the Supplementary Information VI when acquiring the data in Fig. 1 to deterministically obtain the ESR spectral lines corresponding to each nuclear spin state configuration. With initializing nuclei $\mathrm{N_1}$, $\mathrm{N_2}$, and $\mathrm{N_3}$ to $\ket{\Downarrow\Downarrow\Downarrow}$ states, we applied a similar approach to study the ESR spectral lines corresponding to different nuclear spin state configurations of nuclei $\mathrm{N_4}$, $\mathrm{N_5}$, and $\mathrm{N_6}$.

\subsection{NMR spectrum}\label{nmr_spectrum}
\begin{figure*}[h]
    \begin{center}
    \includegraphics[width=1.0\columnwidth]{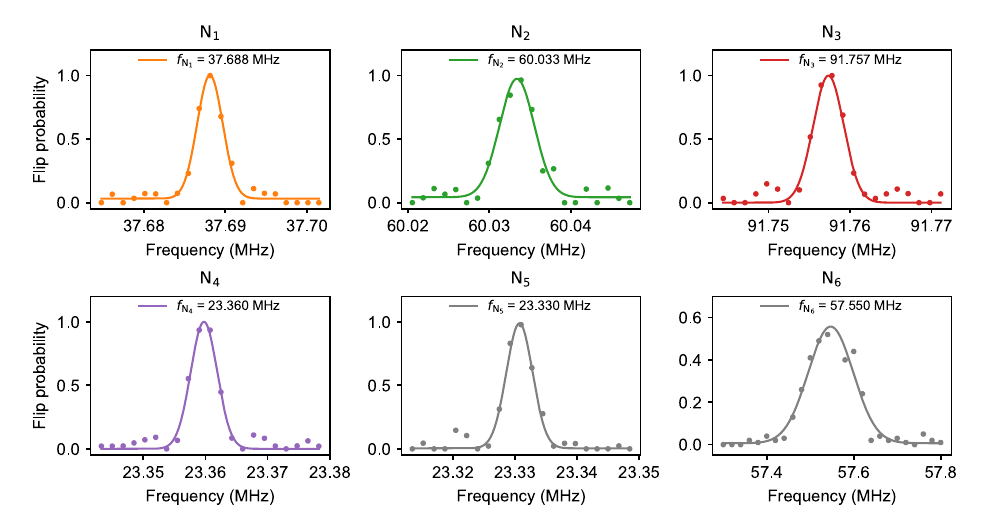}
    \end{center}
    \caption{\textbf{NMR spectra of N$_1$-N$_6$.} All the measured nuclear magnetic resonance spectra are shown, and N$_{1}$-N$_{4}$ are used in the main text.}
    \label{fig:nmr}
\end{figure*}

The NMR frequency can be estimated by $f_\mathrm{N} = \gamma_nB_0 + A/2$, where $\gamma_n$ represents the nuclear gyromagnetic ratio. In experiment, we firstly verify the nuclear spin state by nuclear spin readout, followed by a radio-frequency $\pi$ pulse applied to flip the nuclear spin, and then the resulting nuclear spin state is measured again. The transition events between $\ket{\Uparrow}$ and $\ket{\Downarrow}$ are normalized to obtain the flip probability. The measured NMR spectra of the six nuclei on the electron initialized in $\ket{\downarrow}$ are shown in Fig. \ref{fig:nmr}. For N$_{1}$-N$_{5}$, the experimentally extracted $\gamma_n$ is 17.21, 17.25, 17.34, 17.22, 17.22 $\mathrm{MHz/T}$, respectively, consistent with the gyromagnetic ratio of $^{31}\mathrm{P}$ ($\gamma_n$ = 17.23 $\mathrm{MHz/T}$); and the gyromagnetic ratio for N$_{6}$ is estimated as 42.55 $\mathrm{MHz/T}$, which closely matches the value of $^{1}\mathrm{H}$ ($\gamma_n$ = 42.57 $\mathrm{MHz/T}$).

\clearpage
\subsection{Nuclear spin initialization}\label{nuclear_spin_initialization}
To initialize the multi-spin system (N$_{1}$-N$_{6}$) into a specific spin state combination, such as $\ket{\Downarrow\Downarrow\Downarrow\Downarrow\Downarrow\Downarrow}$, we mainly use the electron state transfer (EST) method, which is realized by a serial of ESR and NMR pulse sequences~\cite{Thorvaldson:2024cww}. Taking N$_{1}$ $\ket{\Downarrow}$ initialization as an example, the electron is firstly initialized to $\ket{\downarrow}$, then we applied ESR $\pi$ pulses at frequencies involving N$_{1}$ being in $\ket{\Downarrow}$ to flip the electron spin; this is followed by the NMR $\pi$ pulse at frequency $v_{\mathrm{n1}}$ to flip N$_{1}$ spin conditional on the electron being in $\ket{\downarrow}$. If N$_{1}$ starts with $\ket{\Downarrow}$, its population will be shelved in $\ket{\uparrow\Downarrow}$ by the electron, thus rendering the pulse at frequency $v_{\mathrm{n1}}$ ineffective; otherwise, N$_{1}$ will be flipped to $\ket{\Downarrow}$ by $v_{\mathrm{n1}}$. Therefore, through this protocol, nuclear N$_{1}$ spin is initialized to $\ket{\Downarrow}$. The same protocol is sequentially applied to the initialization of N$_{2}$, N$_{3}$, N$_{4}$, and N$_{5}$ nuclear spin, achieving a combination of $\ket{\Downarrow\Downarrow\Downarrow\Downarrow\Downarrow}$. The N$_{6}$ nuclear spin that corresponds to $^1\rm{H}$ is periodically checked during the experiment by using QND measurements. When it flips, we will apply an NMR $\pi$ pulse of N$_{6}$ to restore it. Each initialization process contains a QND check step to ensure the initialized state is on the target state. The program will analyze the initialization results in real-time, selecting successful cases through software.

\begin{figure*}[h]
    \begin{center}
    \includegraphics[width=0.6\columnwidth]{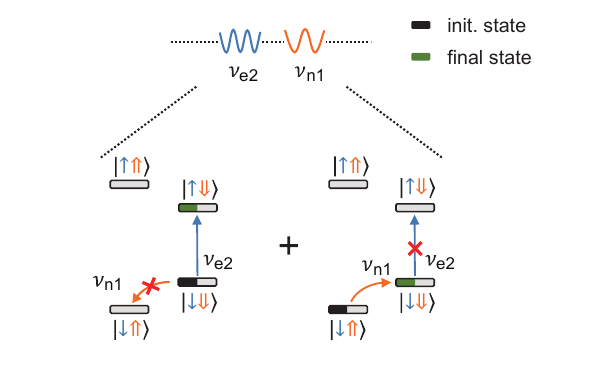}
    \end{center}
    \caption{\textbf{Schematic diagram of nuclear spin initialization.} This is an simplified initialization procedure assuming that N$_2$-N$_6$ are already in their down states. $v_{\mathrm{e2}}$ is the ESR frequency corresponding to nuclear spin configuration of $\ket{\Downarrow\Downarrow\Downarrow\Downarrow\Downarrow\Downarrow}$ and $v_{\mathrm{n1}}$ is the NMR frequency for N$_1$ with electron spin in $\ket{\downarrow}$}
    \label{fig:NSinitial}
\end{figure*}

\clearpage
\subsection{State preparation and measurement error}\label{SPAM}

\begin{figure*}[h]
    \begin{center}
    \includegraphics[width=0.8\columnwidth]{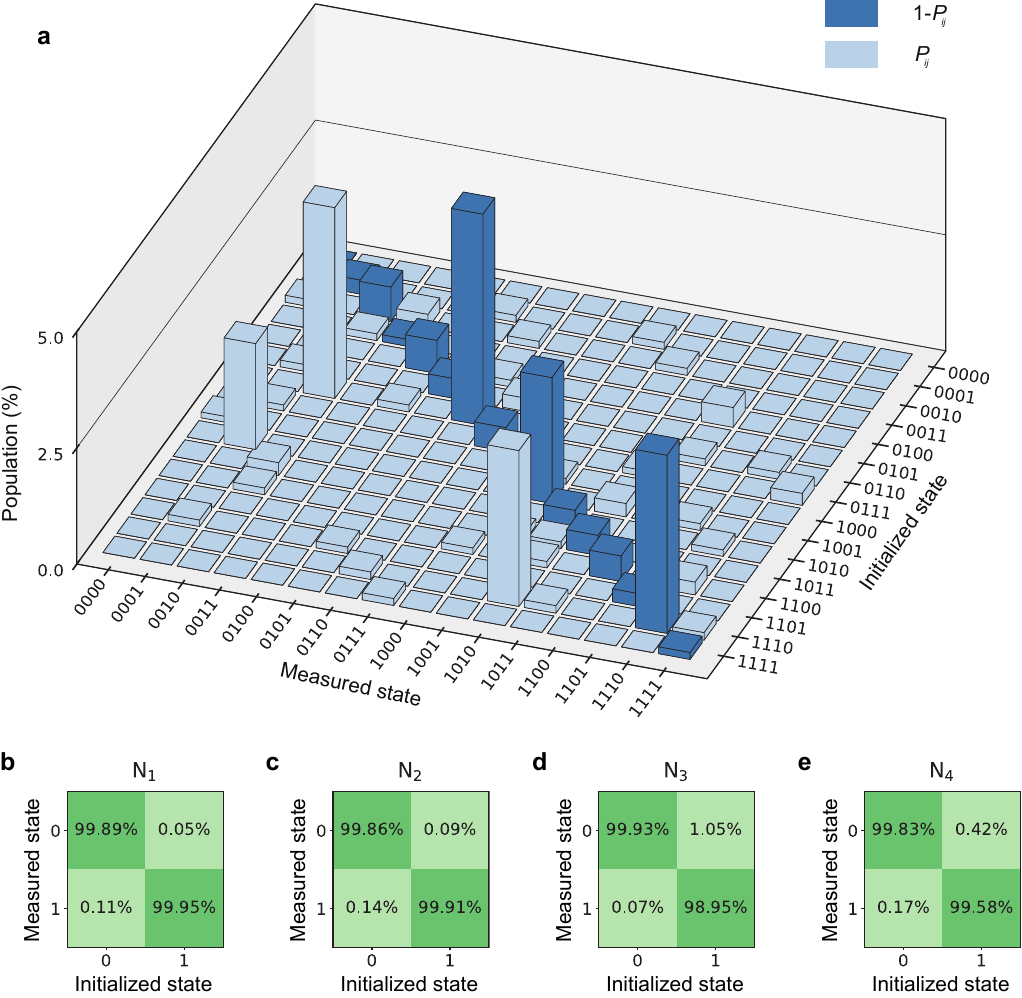}
    \end{center}
    \caption{\textbf{State preparation and measurement error characterization.}
    \textbf{a,} The state preparation and measurement error matrix of the four nuclear spin qubits N$_{1}$-N$_{4}$ utilized in the error detection circuits. The element $P_{ij}$ in the matrix is defined as measured probability for $\ket{i}$ state when initialized in $\ket{j}$ state. Each initialized state is projected across 16 nuclear spin measurement bases, and each is sampled by 1000 shots and then normalized to obtain $P_{ij}$. To highlight the error, (1 - $P_{ij}$) are plotted for the diagonal elements. \textbf{b-e, } The resulting state mapping matrices of each nuclear spin qubit are calculated by marginalization, summing all the probabilities over all states corresponding to the target qubit transition from 0 to 1, while tracing out the other qubits. To quantitatively assess the presence of correlated SPAM errors, we analyzed the two-qubit SPAM matrices for all pairs of qubits. If correlated SPAM errors are negligible, the multi-qubit SPAM matrix can be approximated as the tensor product of individual single-qubit SPAM matrices. Specifically, for each qubit pair $(i,j)$, we obtained the corresponding two-qubit SPAM matrix $M_{ij}$ by tracing out the other two qubits from the full 16×16 experimentally measured SPAM matrix. Similarly, the single-qubit SPAM matrix $M_i$ ($M_j$) was extracted by tracing out the remaining qubits. Using the above method, we constructed the tensor-product approximation $M_{i\otimes j}  = M_i\otimes M_j$ for each qubit pair. We then computed the Frobenius norm of the residual $\| M_{ij}-M_{i\otimes j} \|_{\rm{F}}$, which quantifies the strength of correlated SPAM errors. The computed residuals for all six qubit pairs are consistently small, with values of 0.7\% $(\mathrm{N_4N_3})$, 0.5\% $(\mathrm{N_4N_2})$, 0.6\% $(\mathrm{N_4N_1})$, 1.8\% $(\mathrm{N_3N_2})$, 1.8\% $(\mathrm{N_3N_1})$, and 0.1\% $(\mathrm{N_2N_1})$. These results indicate that while correlated SPAM errors are present, they remain minor across all pairs. We emphasize that SPAM error removal is not applied to any of the data presented in the manuscript.}
    \label{fig:SPAM}
\end{figure*}

\clearpage
\subsection{Nuclear spin quantum jumps caused by ionization shock}\label{nuclear_jumps}
\begin{figure*}[h]
    \begin{center}
    \includegraphics[width=1.0\columnwidth]{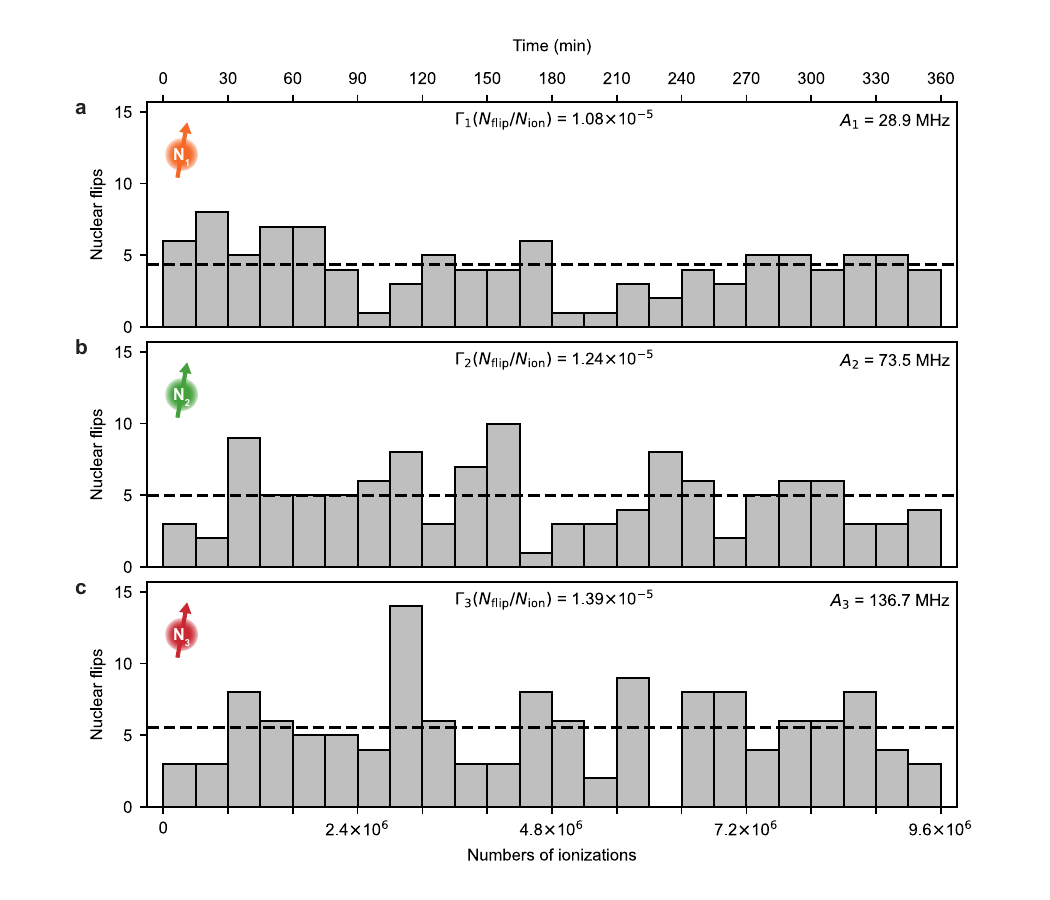}
    \end{center}
    \caption{\textbf{Nuclear spin quantum jumps caused by ionization shock.}
    The nuclear spin readout is realized by electron spin-dependent tunneling between the donor and the SET island. Once the bound electron escapes from the donor, all the hyperfine interactions $A_1$ to $A_6$ vanish. This process is termed "ionization shock", which can induce the nuclear spin flips with a small probability~\cite{mkadzik2022precision}. In this experiment, the electron spin readout (ionization) occurs every 2 ms, and hence, each nuclear spin undergoes 9.6$\times$10$^6$ ionizations over 6 hours. The flip rate $\Gamma$ is defined as the ratio of the nuclear spin flips to the total ionizations throughout the measurement time. \textbf{a,} For nuclear spin qubit N$_{1}$ with $A_1$ = 28.6 MHz, the flip rate is $\Gamma_1 = 1.08\times10^{-5}$. \textbf{b,} For nuclear spin qubit N$_{2}$ with $A_2$ = 73.7 MHz, the flip rate is $\Gamma_2 = 1.24\times10^{-5}$. \textbf{c,} For nuclear spin qubit N$_{3}$ with $A_3$ = 137.0 MHz, the flip rate is $\Gamma_3 = 1.39\times10^{-5}$. The increasing value of $\Gamma$ with hyperfine coupling A indicates that nuclei with larger hyperfine interaction are more susceptible to ionization. Our measured $\Gamma$ is slightly higher than the results in ref.~\cite{mkadzik2022precision}, possibly due to the involvement of more nuclear spins in our cluster. Nonetheless, considering the single-round nuclear spin measurement timescale, the nuclear spin readout can be safely described as a quantum non-demolition measurement.}
    \label{fig:SPAM}
\end{figure*}
\clearpage

\subsection{GST analysis of single-qubit gates}

To identify dominant error sources in nuclear single-qubit gates, we performed the gate set tomography (GST) experiment~\cite{dehollain2016optimization, blume2017demonstration} on N$_1$ as a representative instance.
The GST experiment involves designed sequences composed of fiducial operations and germ sequences. Fiducials are used to prepare and measure a complete set of input and output states, ensuring informational completeness. Germ sequences are short gate patterns specifically chosen to amplify different types of errors. After executing the GST circuits, a MLE method is employed to reconstruct the gate error model that best fits the experimental data.
We implemented the single-qubit GST experiment and analyzed the experimental data using the pyGSTi Python package~\cite{pyGSTi, nielsen2020probing}. 
In our experiment, the state preparation and measurement fiducials are composed of $\{\mathrm{null}, X_{\pi/2}, Y_{\pi/2},X_{\pi/2}X_{\pi/2},X_{\pi/2}X_{\pi/2}X_{\pi/2},Y_{\pi/2}Y_{\pi/2}Y_{\pi/2}\}$ and the germ sequences are constructed from $\{X_{\pi/2}, Y_{\pi/2},X_{\pi/2}Y_{\pi/2},X_{\pi/2}X_{\pi/2}Y_{\pi/2}\} $, with each germ repeated $L$ times, where $ L \in \{1, 2, 4, 8\}$, to enhance sensitivity to small imperfections.
We find that within its single-qubit subspace, gate errors are primarily attributed to Hamiltonian errors, which can be represented as unitary operations, rather than to stochastic errors. This result indicates that the single-qubit gate fidelity of nuclear spin is limited predominantly by control pulse duration imperfections and qubit frequency calibration, rather than by the coherence time.

\begin{figure*}[h]
    \begin{center}
    \includegraphics[width=1\columnwidth]{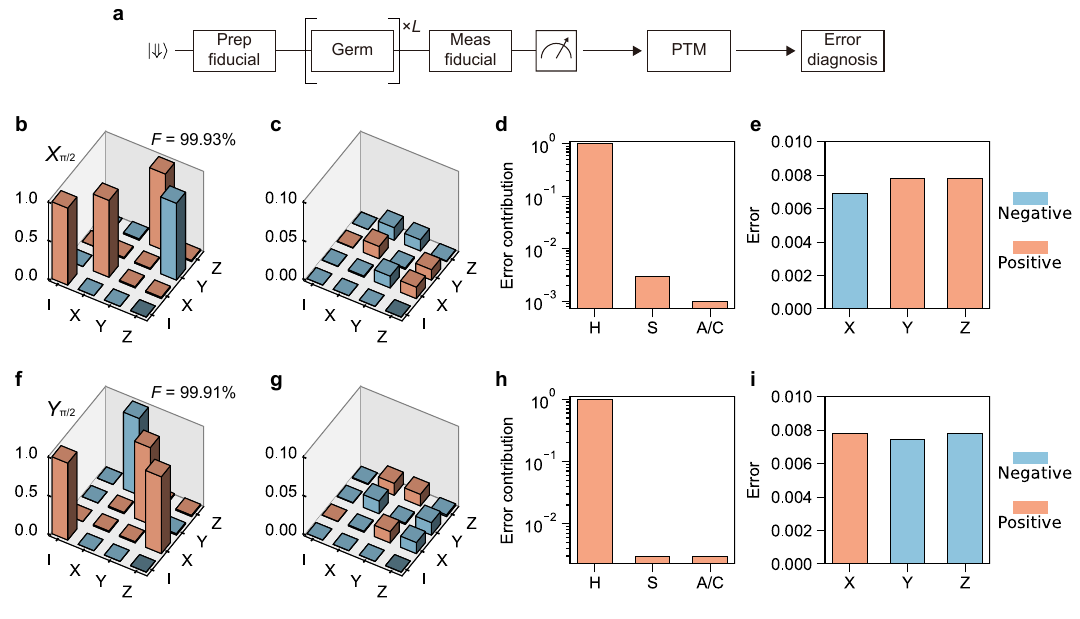}
    \end{center}
    \caption{\textbf{Error analysis of single-qubit gates on qubit N$_1$}. \textbf{a}, Workflow of the GST Experiment. Panels \textbf{b}–\textbf{e} correspond to the GST analysis results for the $X_{\pi/2}$ gate, while panels \textbf{f}–\textbf{i} correspond to the GST analysis results for the $Y_{\pi/2}$ gate. \textbf{b}, \textbf{f}, Pauli transfer matrices (PTMs) extracted from GST experiments. \textbf{c}, \textbf{g}, Error generators $\mathcal{L}$, calculated via $\mathcal{L} = \log(M_{\mathrm{exp}} M_{\mathrm{ideal}}^{-1})$, where $M_{\mathrm{exp}}$ and $M_{\mathrm{ideal}}$ are the experimentally measured and ideal PTMs, respectively. \textbf{d}, \textbf{h}, Decomposition of the gate errors based on the error generator, where H denotes Hamiltonian errors, S denotes stochastic errors, A denotes active errors, and C denotes Pauli correlation errors. \textbf{e}, \textbf{i}, Extracted rotation error components (X, Y, and Z rotations) from the Hamiltonian errors.}
    \label{fig:RB}
\end{figure*}

\clearpage
\subsection{Error analysis of CZ gates and state preparation}\label{error_analysis_CZ_gates}

\begin{figure*}[h!]
			\centering
			\includegraphics[width=0.5\linewidth]{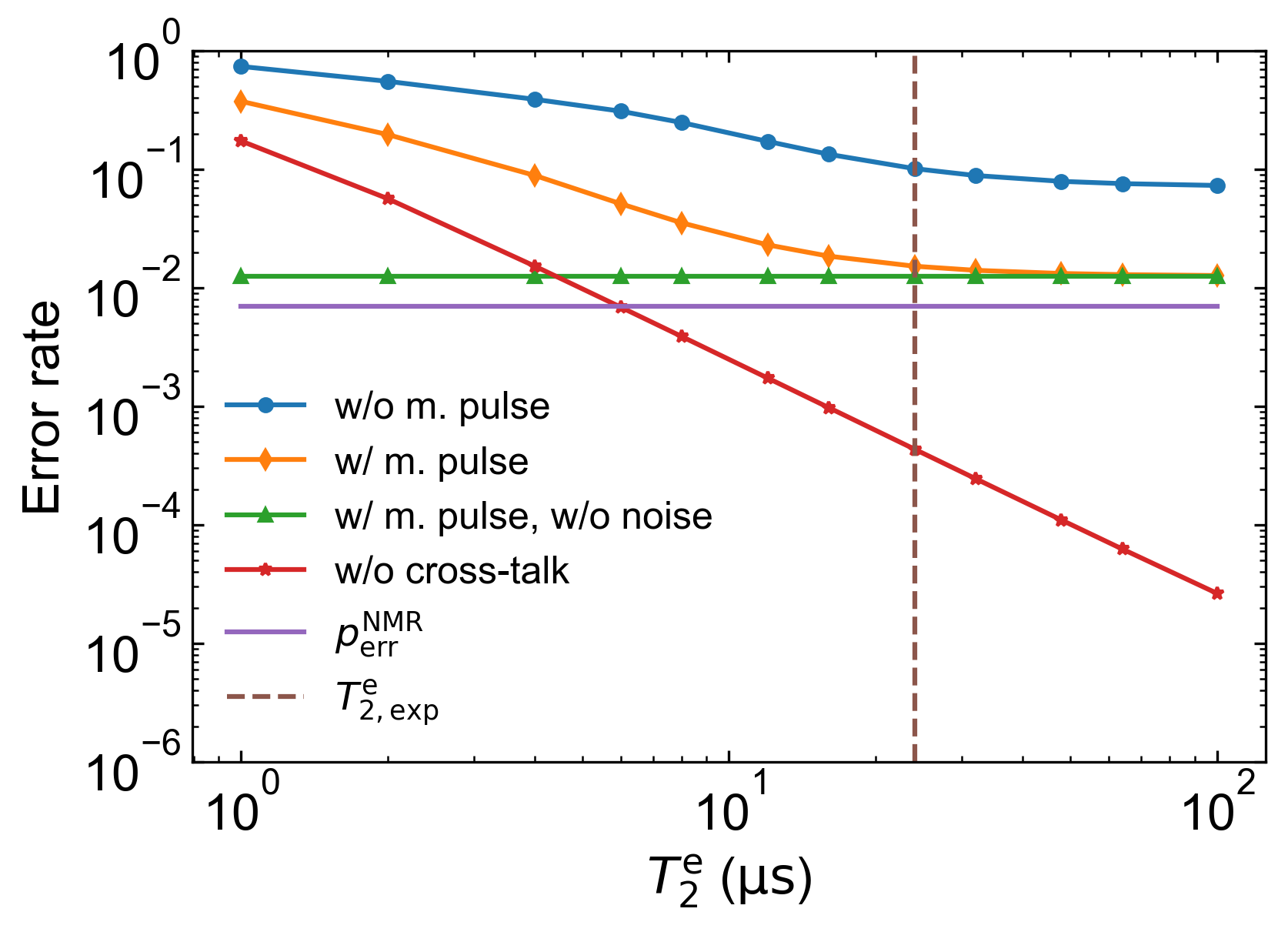}
			\caption{\textbf{Error analysis of CZ gates by simulations.} The simulated error rates of the ESR-type CZ gate are plotted as a function of the electron spin decoherence time under various conditions. The blue line with circles shows the results without mirror pulses (w/o m. pulse) while including the electron spin decoherence noise. In contrast,  the orange lines with diamonds represent the case of applying mirror pulses. The green lines with triangles correspond to the decoherence-free case. To explicitly illustrate the impact of crosstalk error, a red line with stars displays the result without crosstalk error. Additionally, the purple solid line indicates the lowest error rate $p_{\textup{err}}^{\textup{NMR}}$ for NMR-based single-qubit gates, and the brown dashed line marks the electron decoherence time in experiment $T_{\textup{2,exp}}^{\textup{e}}$.}
			\label{fig:error}
		\end{figure*}

The fidelity of the CZ gate drops significantly due to crosstalk, especially when involving nuclear spins with weak hyperfine interactions (HF), as observed for nuclear spin pairs including $\rm{N_4}$ (with $A_4 = 226$ kHz, as shown in Fig. \ref{fig:RB}c). To identify the main error sources, we simulated the CZ gate of $\rm{N_3N_4}$ under various conditions, with the corresponding error rates plotted as a function of electron decoherence time $T_{2}^{\textup{e}}$ in Fig.~\ref{fig:error}. First, the error rate can be suppressed significantly by applying crosstalk mitigation techniques (orange diamonds) introduced in methods, compared to the case without these techniques (blue circles). Under the experimental decoherence conditions (marked by the brown dashed vertical line), the CZ gate error rate shows no significant difference between cases with (orange diamonds) and without (green triangles) the electron spin decoherence noise, confirming that crosstalk errors remain the dominant error source even with mirror pulses. Moreover, the case of removing the crosstalk is also displayed for experimental gate time, i.e., 1$\upmu$s, shown in red line with stars, where the CZ error rate drops below the bare NMR error rates (in purple). In summary, the main methods to maintain a high fidelity of such CZ gate are to avoid using the nuclear spins with weak HF (via proper initialization) or to implement proper crosstalk mitigation techniques.

\begin{figure*}[h!]
			\centering
			\includegraphics[width=0.7\linewidth]{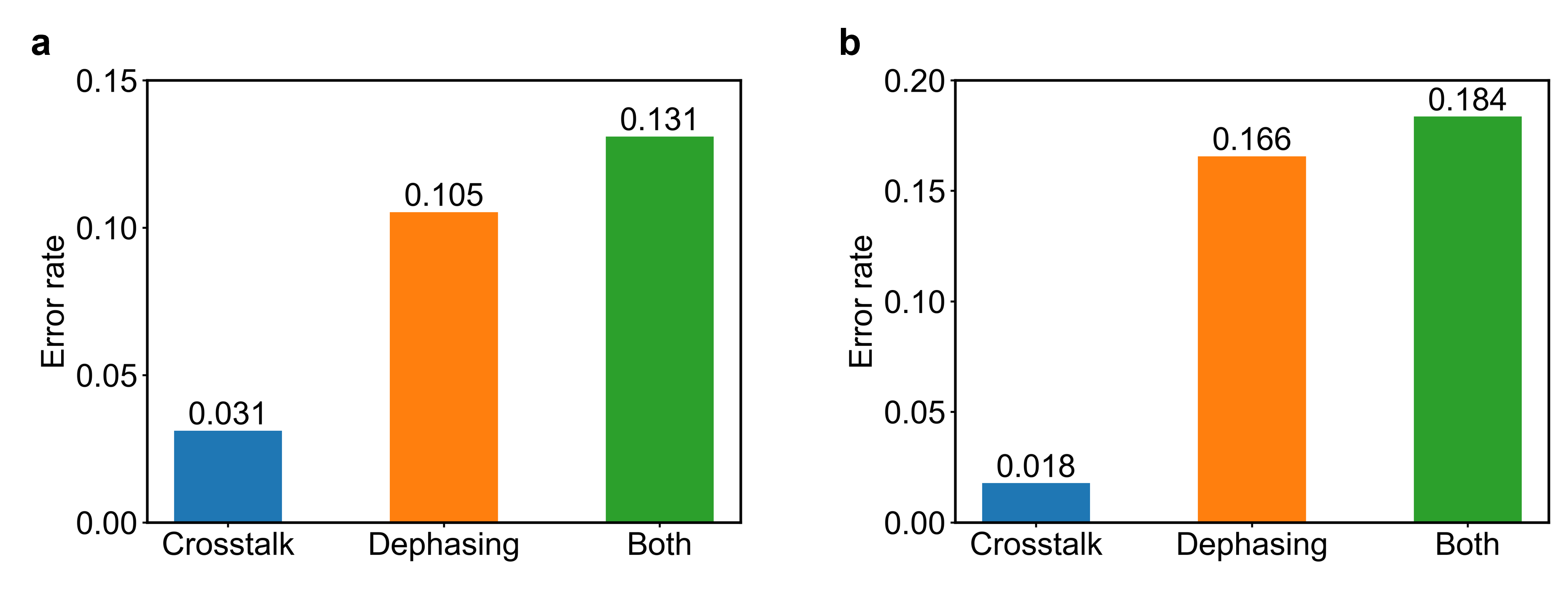}
			\caption{\nonstopmode\textbf{Error analysis for the GHZ state and the recovered Bell state from the equal-superposition state of the four Bell codewords.} The error rate represents the infidelity of the target state under different conditions. The first bar corresponds to the simulation result considering only crosstalk error, the second bar represents the simulation result with only dephasing-induced error, and the third bar shows the simulation result with both crosstalk error and dephasing-induced error. \textbf{a}, Error analysis for the GHZ state. \textbf{b}, Error analysis for the recovered Bell state from the superposition state of four Bell codewords.}
			\label{fig:error_GHZ}
		\end{figure*}

Next, the error types in entangled state preparation are analyzed. While crosstalk is the dominant error source for ESR-based CZ gates, the state preparation circuit involves NMR-based manipulations that are primarily affected by qubit dephasing. To quantify the relative contributions of these error sources in specific circuits, master equation simulations were performed under three distinct conditions: only crosstalk errors, only dephasing-induced errors, and both error sources considered simultaneously. In the following discussion, error rate refers to the infidelity when only the corresponding error sources are considered. The results for GHZ state preparation are shown in Fig.~\ref{fig:error_GHZ}a. The error rate induced by crosstalk (3.1\%) is lower than that induced by dephasing (10.5\%). Therefore, for GHZ state preparation, qubit dephasing is the dominant error source. Results for the Bell state recovered from the equal-superposition Bell state (see Extended Data Fig. 7) led to the same conclusion. As shown in Fig.~\ref{fig:error_GHZ}b, the crosstalk error rate (1.8\%) is much lower than the dephasing-induced error rate (16.8\%).

\begin{figure*}[h!]
			\centering
			\includegraphics[width=1\linewidth]{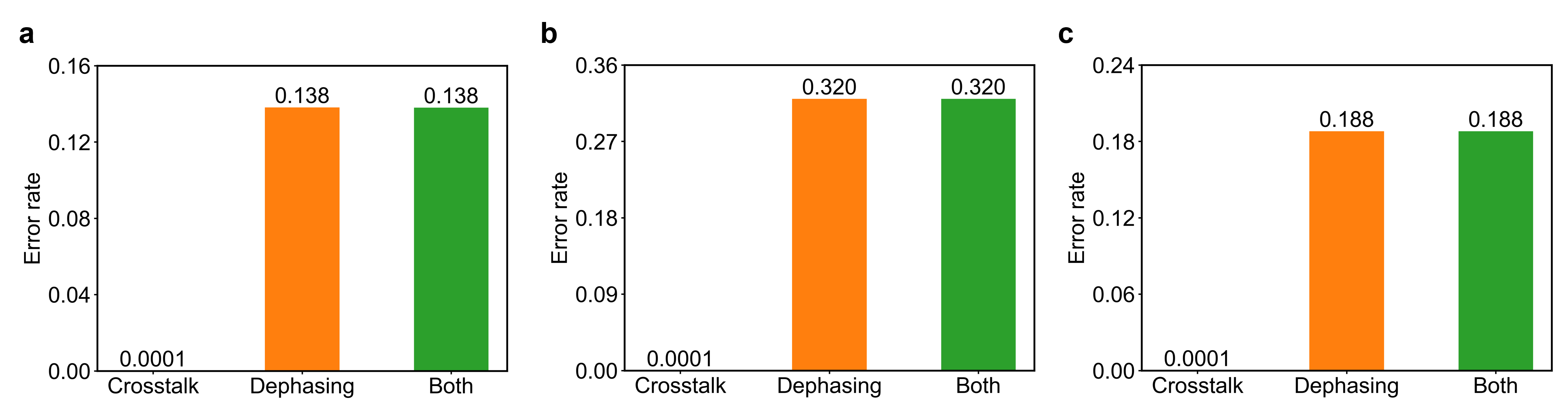}
			\caption{\nonstopmode\textbf{Error analysis for the initial Bell state, the uncorrected Bell state, and the recovered Bell state.} The error rate represents the infidelity of the target state under different conditions. The first bar corresponds to the simulation result considering only crosstalk error, the second bar represents the simulation result with only dephasing-induced error, and the third bar shows the simulation result with both crosstalk error and dephasing-induced error considered. \textbf{a,} Error analysis for the initial Bell state with $t_\textup{wait}=0$. \textbf{b,} Error analysis for the uncorrected Bell state with $t_\textup{wait}=192$ $\upmu$s. \textbf{c,} Error analysis for the recovered Bell state.}
			\label{fig:error_Bell}
		\end{figure*}

Simulation results for Bell state $\ket{\Phi^+}$ preparation circuit with the error detection circuit are shown in Fig.~\ref{fig:error_Bell}. The circuit exhibits a low crosstalk error rate (0.01\%), which confirms dephasing as the primary error source for the initial, uncorrected, and recovered Bell state. For the initial Bell state with $t_\textup{wait}=0$, the dephasing-induced error rate is 13.8\% (Fig.~\ref{fig:error_Bell}a). With the waiting time $t_\textup{wait}=192$ $\upmu$s, the dephasing-induced error rate increases to 32\% (Fig.~\ref{fig:error_Bell}b). Then, through post-processing based on error detection results, the error rate is reduced to 18.8\% (Fig.~\ref{fig:error_Bell}c). When considering both types of error, the error rates vary similarly to the case with only dephasing-induced errors.

Notably, even after post-processing, the Bell state can only be partially recovered. This occurs primarily because the error detection circuit itself introduces errors that cannot be detected and recovered. The discrepancy in the recovered Bell state between simulation and experimental results is likely to be due to differences between simulated and practical noise, as well as unaccounted control and SPAM error in the simulation. In summary, these results highlight that the decoherence of the spin qubit system is the main limiting factor for the state fidelities, and demonstrate the importance of quantum error correction for large-scale quantum computation.
        
\clearpage
\subsection{Biased error}\label{biased_error}
\begin{figure*}[h!]
			\centering
			\includegraphics[width=0.4\linewidth]{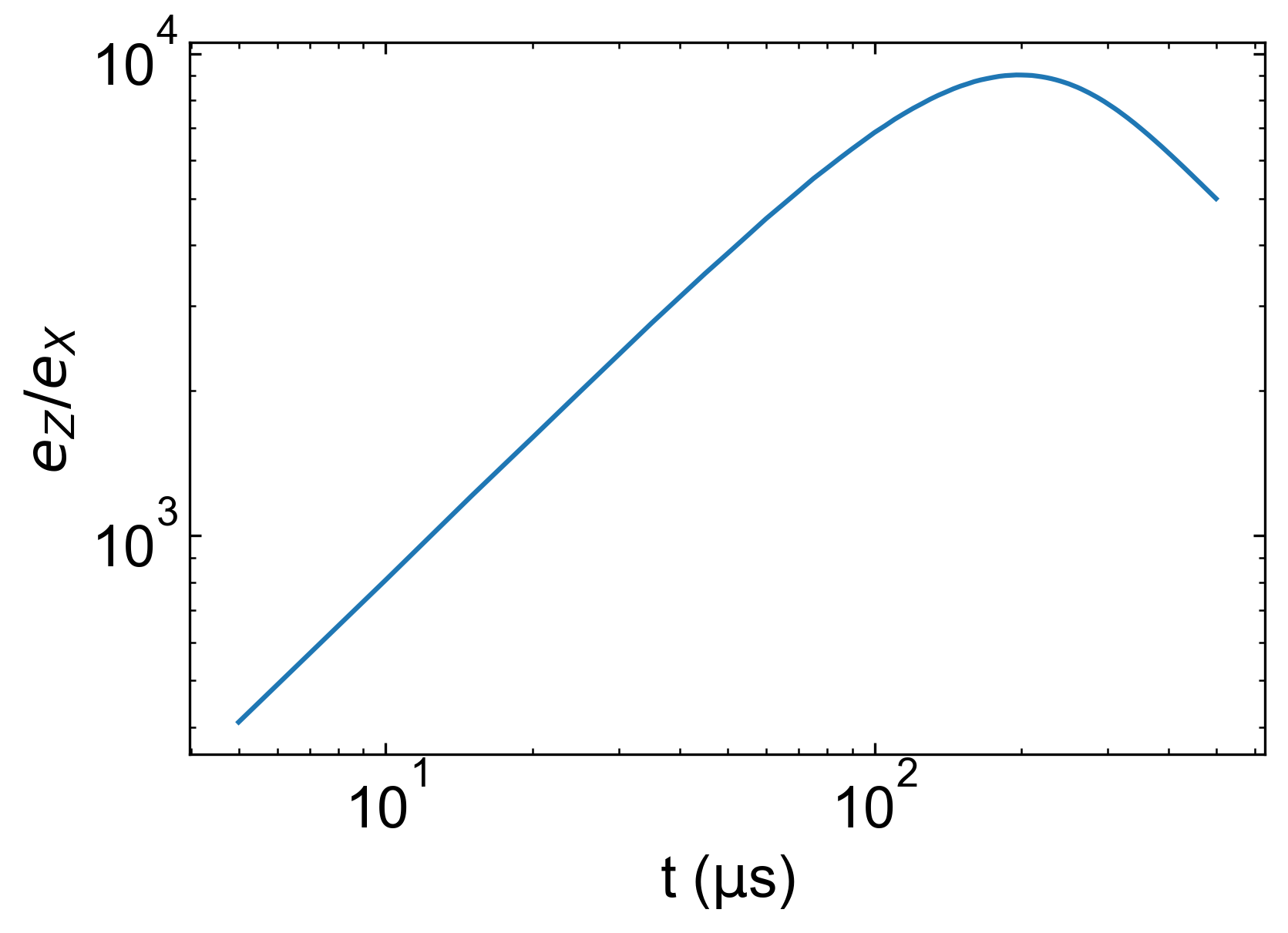}
			\caption{\nonstopmode\textbf{The ratio of Z to X errors on a Bell state as a function of time.} The relaxation time ($T_1$) of the qubits constituting the Bell state is $10^4$ times longer than their dephasing time ($T_2^*$).}
			\label{fig:zx_ratio}
		\end{figure*}

Facilitated with biased error, tailored quantum error correction architectures allow a higher fault-tolerance threshold and can reduce the overhead for physical qubit numbers per logical qubit~\cite{roffe2023biastailoredquantum,wu2025bias}. For spin qubits in silicon, the relaxation time ($T_{1}$) is significantly longer than the dephasing time ($T_2^*$)~\cite{Madzik2021,Tanttu2024-nu}. In the case of P-donor nuclear spin qubits, the dephasing time $T_2^*$ is approximately on the order of milliseconds~\cite{muhonen2014storing,edlbauer2025}, while the relaxation time $T_1$ exceeds the order of seconds~\cite{Pla2013}. These observations suggest the presence of strongly biased errors. The ratio of Z errors ($e_Z$) to X errors ($e_X$) is proportional to the ratio of $T_1$ to $T_2^*$, with the upper bound of the bias ratio of errors $e_Z /e_X$ being determined by $T_1 /T_2^*$~\cite{martinez2020approximating,hetenyi2024bias}. To quantitatively verify the significant error bias ratio under a high $T_1/T_2^*$ ratio, we have numerically simulated the time evolution of a two-qubit Bell state with $T_1/T_2^*=10^4$. As shown in Fig.~\ref{fig:zx_ratio}, during the 500 $\upmu$s evolution, the error ratio $e_Z/e_X$ remains within the range of $10^2$ to $10^4$, illustrating strong biased error.

An additional potential source of X/Y errors arises from crosstalk due to the weak hyperfine coupling of $\mathrm{N_4}$. In this work, the crosstalk errors are effectively suppressed by crosstalk mitigation techniques (see Methods). Master equation simulations on the Bell state preparation and error detection circuit (see Supplementary Information, Sec.~\ref{error_analysis_CZ_gates}) reveal that the crosstalk error rate (0.01\%) is much lower than the dephasing-induced error rate (13.8\%). With future optimizations to reduce the frequency crowding issue in cluster-based devices, the bias-tailored quantum error correction scheme can be implemented in P-donor spin qubit systems.
\clearpage

\subsection{Scalable cluster-based qubit scheme}\label{scalable_scheme}
In this section, we discuss the scalability issue for the donor-cluster system. We propose utilizing the donor clusters for scalable quantum computing, in which the nuclear spins and the electron spins are encoded as data qubits and ancillary qubits, respectively.
Although the device in the main text demonstrates a small-scale qubit platform, this strategy is not inherently scalable due to frequency crowding when more nuclei are added. The key challenge is to scale up the number of qubits without suffering from the frequency crowding problem.

To determine the maximum feasible number of donors within a single cluster, we have conducted numerical simulations by randomly sampling donors with hyperfine interaction strengths between the minimum values of 0.6 MHz~\cite{edlbauer2025} and maximum values of 304 MHz~\cite{kranz2023use_of_exchange} reported from experimental devices. The sampling procedure continues until the minimum difference between any two selected hyperfine interactions fell below the 10 MHz threshold, which maintains crosstalk errors below 1\% at an electron Rabi frequency of 0.5 MHz. Across thousands of simulations, the maximum feasible number of donors reaches 11, while the minimum is only 2. On average, the number of selectable donors is 4.3 (1.5). With crosstalk mitigation techniques (see Methods) or a slower ESR drive, the averaged donor number can be increased further. However, these improvements cannot overcome the inherent scalability limitations of single-cluster architectures for large-scale quantum computing. To address the scalability problem, we leverage the advantage of electron spins, as demonstrated in gate-defined quantum systems using scalable quantum dot arrays~\cite{RevModPhys.95.025003}, and propose a sheme for scalable donor-based quantum computing. 

\begin{figure*}[hbt!]
			\begin{center}
				\includegraphics[width=0.4\columnwidth]{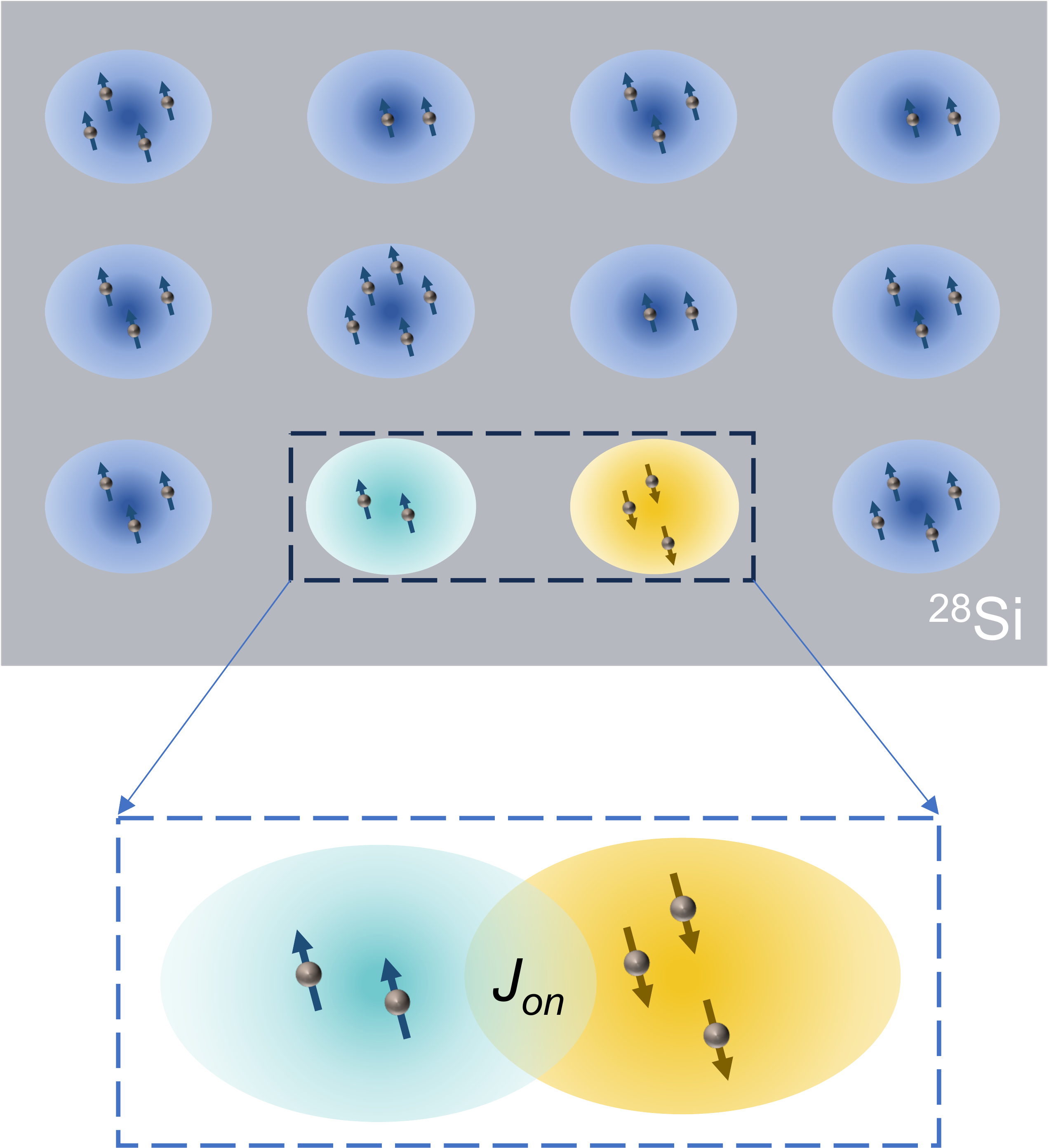}
			\end{center}
			\caption{\textbf{Scheme for nuclear-spin based scalable quantum computing in silicon} The schematic of the two-dimensional cluster array is presented. The lower panel depicts a pair of adjacent clusters, containing two and three P-donors, respectively, and the exchange coupling $J_{on}$ is turned on between the electrons in the pair of clusters with a single electron in each cluster.
            }
			\label{fig:scalable}
		\end{figure*}

The proposed scheme for scalable donor-based quantum computing utilizes the cluster composed of several donors as a building block, forming a two-dimensional array, as shown in Fig.~\ref{fig:scalable}, different from the scheme based on single donors~\cite{stemp2025scalable,Madzik2021}. The intra-cluster entanglement between nuclear spin qubits is still mediated by the shared bond electron. Whereas, the inter-cluster entanglement between nuclear spin qubits is mediated by the Heisenberg exchange interaction between bound electrons.
Here, the shared electron not only facilitates addressability for nuclear spin qubits but also mediates nuclear two-qubit entanglement. Note that the high tunability of the exchange coupling can be achieved by modulating detuning~\cite{He2019,Voisin2020exchange}  or by using the superexchange coupling~\cite{Srinivasa2015superexchange,Baart2017superexchange,zhang2024superexchange}. Once tunable coupling is available, the exchange interaction with electrons in the nonrelevant clusters can be turned off, and since the number of nuclear spins can be small in one pair of coupled clusters, we can avoid the frequency-crowding problem that is prevailing in systems with large numbers of coupled spins. Thus, given that the exchange coupling between the electrons can in principle be turned on and off to mediate nuclear spin entanglement and avoid the frequency crowding issue, the donor-cluster system is promising for scalable quantum computing. For more details, please refer to our upcoming theoretical proposal for scalable quantum computing based on donor-cluster system.

\clearpage

\subsection{Calibration circuit of the CCCZ gate}\label{cccz_calibration}

\begin{figure*}[h]
    \begin{center}
    \includegraphics[width=1\columnwidth]{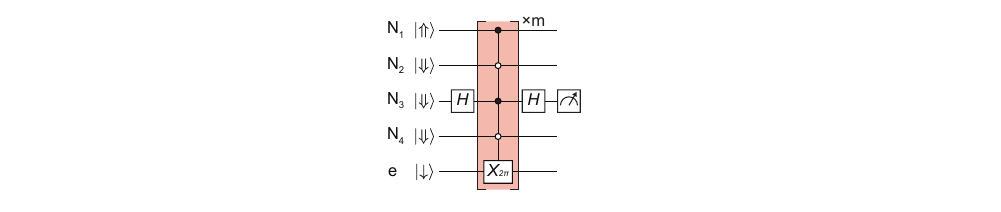}
    \end{center}
    \caption{\textbf{CCCZ gate calibration circuit.} The calibration on single-qubit gates of electron spin and nuclear spins is trivial, and simple Rabi flops can almost give an ideal result. This figure presents an example of the calibration circuit for one of the CCCZ gates. By combining two $\pi/2$ NMR pulses, the control phase on the target state is converted into the flipping probability of the nuclear spin. Here, we measure the spin-up probability of N$_3$ as a function of the electron ESR pulse duration. The resulting data curve is then fitted with a sine function, where the pulse duration corresponding to the maximum value represents the time required to achieve a $2\pi$ rotation of the electron spin. To enhance the sharpness of the interference pattern and thereby improve the operational precision, multiple ESR pulses are applied. In our experiment, the number of pulse repetitions $m$ is increased sequentially as $m = 1$, 3, 5, and 7 to accurately determine the ESR pulse duration for the CCCZ gate.}
    \label{fig:stabilizer}
\end{figure*}

\clearpage
\subsection{The compiled error detection circuit}\label{Thecompiled_error_detection_circuits}
Decoherence and ESR frequency crosstalk are balanced carefully for the circuit compilation. (1) $\mathrm{N_2}$ and $\mathrm{N_3}$ are selected for Bell state encoding is mainly due to their single-qubit gates being the fastest among all nuclear spin qubits, for which the impact of dephasing can be reduced as much as possible during the encoding stage. (2) $\mathrm{N_1}$ is selected as the $\hat{S}^{X}$ stabilizer for phase-flip error detection (marked in purple), $\mathrm{N_4}$ is selected as the $\hat{S}^{Z}$ stabilizer for bit-flip error detection (marked in blue), and the circuit executes stabilizer measurement on $\mathrm{N_1}$ before $\mathrm{N_4}$. The phase-flip error is the main error source in our biased system, so we first perform $\hat{S}^{X}$ stabilizer measurement using $\mathrm{N_1}$ with a relatively fast single-qubit gate and little ESR frequency crosstalk. On the other hand, $\mathrm{N_4}$ exhibits not only the slowest driving speed but also the smallest hyperfine interaction $\sim$226~kHz. Once it is in the superposition state, the ESR gate time has to be prolonged to reduce the crosstalk. As a result, we perform $\hat{S}^{Z}$ stabilizer measurement using $\mathrm{N_4}$ at the end of the circuit to ensure the circuit performance.
\begin{figure*}[h]
    \begin{center}
    \includegraphics[width=1.0\columnwidth]{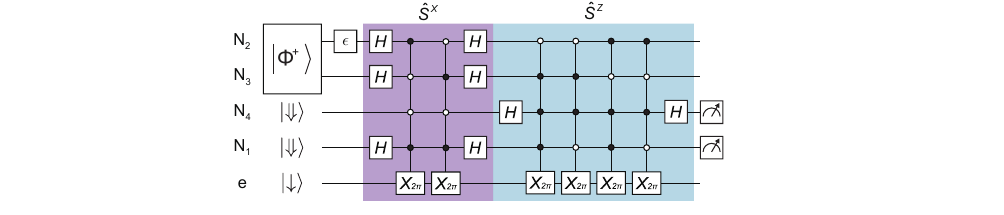}
    \end{center}
    \caption{\textbf{The compiled error detection circuit.} Since our native gates consist of single-qubit gates on the nuclear spins and ESR type CCCZ gates, we compile the error detection circuits shown in Fig. 3a using these native gates. In the complied circuit, the fifth qubit represents the electron, the purple-shaded region indicating the $\hat{S}^{X}$ stabilizer, and the blue-shaded region representing the $\hat{S}^{Z}$ stabilizer.}
    \label{fig:stabilizer}
\end{figure*}

\begin{figure*}[h]
    \begin{center}
    \includegraphics[width=1.0\columnwidth]{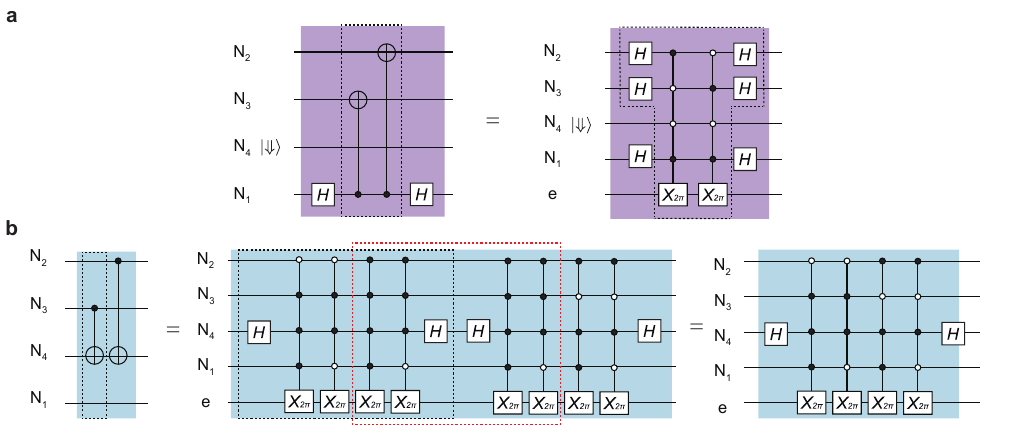}
    \end{center}
    \caption{\nonstopmode\textbf{Error detection circuit decomposition.} \textbf{a,} An equivalent circuit decomposition into CCCZ type gates and Hadamard gates for the stabilizer $\hat{S}^{X}$. Since N$_4$ is in the spin-down state, we do not need to apply the ESR pulse corresponding to N$_4$ being in the spin-up state. Therefore, the operations within the black dashed box are effectively equivalent. \textbf{b,} Circuit decomposition for the stabilizer $\hat{S}^{Z}$. Each CNOT gate is decomposed into four ESR pulses and two Hadamard gates. The operations within the black dashed box are effectively equivalent, while those enclosed by the red dashed box effectively constitute an identity operation and can therefore be omitted, resulting in the final compiled circuit for the $\hat{S}^{Z}$ stabilizer.}
    \label{fig:circuit_decomposition}
\end{figure*}

\clearpage
\subsection{Crosstalk of N$_4$ and N$_5$}\label{N4N5}
To verify that there is no crosstalk and leakage error between nuclear spin qubits $\mathrm{N_4}$ and $\mathrm{N_5}$, we conducted the following experiment. As shown in Fig. S18a, $\mathrm{N_4}$ and $\mathrm{N_5}$ are initialized to a fixed state. We then applied a $\pi$/2 rotation to $\mathrm{N_4}$ with increasing input power while monitoring the populations of both qubits (Fig. S18b). The population of $\mathrm{N_4}$ remained within its original state space across all power levels, manifesting no leakage error. However, the population of $\mathrm{N_5}$ started to rise at an input power above 17 dBm, which originates from the NMR crosstalk. Therefore, we conclude that there is little crosstalk error under the driving power for $\mathrm{N_4}$ (black dotted line) in our experiment.
\\
\begin{figure*}[h]
    \vspace{0.50cm}
    \begin{center}
    \includegraphics[width=1.0\columnwidth]{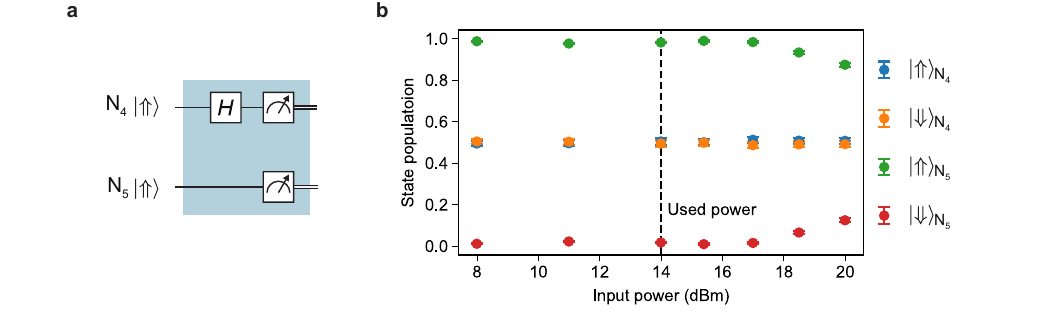}
    \end{center}
    \vspace{-0.60cm}
    \caption{\nonstopmode\textbf{$\mathbf{N_4N_5}$ crosstalk measurement.} \textbf{a,} The circuit utilized to measure the crosstalk between $\mathrm{N_4}$ and $\mathrm{N_5}$. \textbf{b,} $\mathrm{N_4N_5}$ state population with different $\mathrm{N_4}$ driving power, the black dashed line indicates the power used to drive $\mathrm{N_4}$ in the experiment of the main text.}
    \label{fig:circuit_decomposition}
\end{figure*}

In the future, we may use some crosstalk mitigation method for NMR operation, which requires setting drive power for the $\mathrm{N_4}$ $\pi$/2 rotation to simultaneously introduce a 2$\pi$ rotation on $\mathrm{N_5}$ under the non-resonant driving conditions. The drive strength $f_D$(in Hz) should satisfy:
\begin{equation}
    \frac{\pi/2}{f_D} = \frac{k\cdot 2\pi}{\sqrt{f_D^2 + (f_{\mathrm{N}_4} - f_{\mathrm{N}_5})^2}}  
\end{equation}

where $k$ is an integer ($k$ = 1, 2, 3, ...).

\clearpage
\renewcommand{\tablename}{Supplementary Table.}
\subsection{Supplementary tables}
\subsubsection{Qubit parameters}
\begin{table*}[h]
    \centering
    
    \begin{tabular}{c|c|c|c|c}
    
    \hline
    
    \   & $\omega_q/2\pi$ [GHz] & $T^*_2$[$\upmu$s]  & $\pi$/2 gate duration [$\upmu$s]   & RB [$\%$] \\
    
    \hline
    
    $e_{\Uparrow\Uparrow\Uparrow\Uparrow\Uparrow\Uparrow}$
    & 37.825324 & 23.4(0.5) & 0.30 & 95.55(0.06)  \\
    \hline
    $N_{1\downarrow}$& 0.03768 & 441(11) & 25.74 & 99.37(0.03)  \\
    $N_{2\downarrow}$& 0.06003 & 349(8) & 16.20 & 99.73(0.02) \\
    $N_{3\downarrow}$& 0.09175 & 788(23) & 16.20 & 99.84(0.01) \\
    $N_{4\downarrow}$& 0.02336 & 24.8(0.8)$\times 10^3$ & 40.47 & 99.33(0.04)  \\
    
    \hline
    \end{tabular}
    
    \caption{\textbf{Qubit parameters.} \raggedright The qubit frequency $\omega_q/2\pi$, $T^*_2$, gate duration of $X_{\pi/2}$ pulse, and single-qubit RB fidelities are listed for the electron and four encoding nuclei spins in this work.}
    \label{table:Qubit_parameters}
\end{table*}
\subsubsection{Operations applied for restoring the encoded entanglement information.}
\begin{table*}[h]
    \centering
    \begin{tabular}{cccccc}
		\toprule
		$\rho_{\rm{input}}$ 
		& \{$M_1$,$M_2$\} & Error & $\rho_{\rm{error}}$ & $U$ & 
		$\rho_{\text{corrected}}$\\
		\midrule
		\multirow{4}{*}{$\Ket{\Phi^+}$} 	
        & \{0,0\} & None & $\Ket{\Phi^+}$ & $II$ & \multirow{4}{*}{$\Ket{\Phi^+}$}\\
        \addlinespace[1.5 pt]
		& \{0,1\} & Z    & $\Ket{\Phi^-}$ & $IZ$            & \\
        \addlinespace[1.5 pt]
		& \{1,0\} & X    & $\Ket{\Psi^+}$ & $IX$            & \\
        \addlinespace[1.5 pt]
		& \{1,1\} & Y    & $\Ket{\Psi^-}$ & $IY$            & \\
		\addlinespace[1.5 pt]
        \hline
        \addlinespace[1.5 pt]
		\multirow{4}{*}{$\Ket{\Psi^-}$} 	
		& \{0,0\} & Y    & $\Ket{\Phi^+}$ & $IY$            & \multirow{4}{*}{$\Ket{\Psi^-}$}\\
        \addlinespace[1.5 pt]
		& \{0,1\} & X    & $\Ket{\Phi^-}$ & $IX$            & \\
        \addlinespace[1.5 pt]
		& \{1,0\} & Z    & $\Ket{\Psi^+}$ & $IZ$            & \\
        \addlinespace[1.5 pt]
		& \{1,1\} & None & $\Ket{\Psi^-}$ & $II$ & \\
		\bottomrule
     
    \end{tabular}
    \caption{\textbf{Operations applied for restoring the encoded entanglement information.} Here we provide some details on implementing error detection and entanglement restoration for the Bell states by postprocessing. There are four maximally entangled Bell states $\ket{\Phi^+} = \frac{1}{\sqrt{2}}(\ket{00}+\ket{11})$, $\ket{\Phi^-} = \frac{1}{\sqrt{2}}(\ket{00}-\ket{11})$, $\ket{\Psi^+} = \frac{1}{\sqrt{2}}(\ket{01}+\ket{01})$, and $\ket{\Psi^-} = \frac{1}{\sqrt{2}}(\ket{01}-\ket{10})$. Any single-qubit error will transform a prepared Bell state into one of the other three. For example, a Z error will transform $\ket{\Phi^+}$ to $\ket{\Phi^-}$. By performing parity measurements of the XX and ZZ stabilizers, different Bell states produce distinct measurement outcomes regarding the syndrome qubit. This allows the detection of errors that may have occurred in the prepared Bell state $\rho_{\rm{input}}$. As shown in the table, the cases for $\ket{\Phi^+}$ and $\ket{\Psi^-}$ as the input states, along with corresponding operations are summarized. For a given input state, different syndromes are related to four different error states, each associated with unique unitary operations that can recover the input state.}
	\label{Tab}	
\end{table*}
\stoplist[local]{toc}

\clearpage

\bibliography{main.bbl}
\end{document}